\documentstyle[12pt,epsf]{article}
\newlength{\pubnumber} 

\catcode`\@=11
\@addtoreset{equation}{section}

\def\section{\@startsection{section}{1}{\z@}{3.5ex plus 1ex minus .2ex}
 {2.3ex plus .2ex}{\large\bf}}
\def\subsection{\@startsection{subsection}{2}{\z@}{2.3ex plus .2ex}
 {2.3ex plus .2ex}{\bf}}

\begin{document}

\begin{titlepage}
\samepage{
\setcounter{page}{1}
%\rightline{Preliminary DRAFT}
\rightline{IASSNS--HEP--96--44}
\rightline{UFIFT--HEP--96--12}
\rightline{\tt hep-ph/9605325}
\rightline{May 1996}
\vfill
\begin{center}
 {\Large \bf  Stable Superstring Relics\\}
\vfill
\vfill
 {\large Sanghyeon Chang$^{1}$\footnote{
   	E-mail address: schang@phys.ufl.edu},
	Claudio Corian\`{o}$^{1}$\footnote{
   	E-mail address: coriano@phys.ufl.edu}
   	$\,$and$\,$ Alon E. Faraggi$^{1,2}$\footnote{
   	E-mail address: faraggi@phys.ufl.edu}\\}
\vspace{.12in}
 {\it $^{1}$   Institute for Fundamental Theory, Department of Physics, \\
	University of Florida, Gainesville, FL 32611,
	USA\footnote{Permanent address.}\\}
\vspace{.075in}
 {\it  $^{2}$ School of Natural Sciences, Institute for Advanced Study\\
  Olden Lane, Princeton, NJ 08540 USA\\}
\end{center}
\vfill
\begin{abstract}
  {\rm We investigate the cosmological constraints on exotic stable matter
states which arise in realistic free fermionic superstring models.
These states appear in the superstring models due to a ``Wilson--line''
breaking of the unifying non--Abelian gauge symmetry.
In the models that we consider the unifying  $SO(10)$
gauge symmetry is broken at the string level
to $SO(6)\times SO(4)$, $SU(5)\times U(1)$ or
$SU(3)\times SU(2)\times U(1)^2$.
The exotic matter states are classified
according to the patterns of the $SO(10)$ symmetry breaking.
In $SO(6)\times SO(4)$ and $SU(5)\times U(1)$ type models
one obtains fractionally charged states with $Q_{\rm e.m.}=\pm1/2$.
In $SU(3)\times SU(2)\times U(1)^2$ type models
one also obtains
states with the regular charges under the Standard Model
gauge group but with ``fractional'' charges under the $U(1)_{Z^\prime}$
symmetry. These states include down--like color triplets
and electroweak doublets, as well as states which are
Standard Model singlets. By analyzing the renormalizable
and nonrenormalizable terms of the superpotential in a specific superstring
model, we show that these exotic states can be stable.
We investigate the cosmological constraints on the masses and
relic density of the exotic states. We propose that, while
the abundance and the masses of the fractionally charged states
are highly constrained, the Standard Model -- like states,
and in particular the Standard Model singlet,
are good dark matter candidates.
		 }
\end{abstract}
\smallskip}
\end{titlepage}

\setcounter{footnote}{0}

% ========================= DEFINITIONS ===================================
\def\beq{\begin{equation}}
\def\eeq{\end{equation}}
\def\beqn{\begin{eqnarray}}
\def\eeqn{\end{eqnarray}}

\def\ie{{\it i.e.}}
\def\eg{{\it e.g.}}
\def\half{{\textstyle{1\over 2}}}
\def\third{{\textstyle {1\over3}}}
\def\quarter{{\textstyle {1\over4}}}
\def\m{{\tt -}}
\def\p{{\tt +}}

\def\slash#1{#1\hskip-6pt/\hskip6pt}
\def\slk{\slash{k}}
\def\GeV{\,{\rm GeV}}
\def\TeV{\,{\rm TeV}}
\def\y{\,{\rm y}}
\def\SM{Standard-Model }
\def\SUSY{supersymmetry }
\def\SSSM{supersymmetric standard model}
\def\vev#1{\left\langle #1\right\rangle}
\def\l{\langle}
\def\r{\rangle}

\def\Htw{{\tilde H}}
\def\chibar{{\overline{\chi}}}
\def\qbar{{\overline{q}}}
\def\ibar{{\overline{\imath}}}
\def\jbar{{\overline{\jmath}}}
\def\Hbar{{\overline{H}}}
\def\Qbar{{\overline{Q}}}
\def\abar{{\overline{a}}}
\def\alphabar{{\overline{\alpha}}}
\def\betabar{{\overline{\beta}}}
\def\tautwo{{ \tau_2 }}
\def\thetatwo{{ \vartheta_2 }}
\def\thetathree{{ \vartheta_3 }}
\def\thetafour{{ \vartheta_4 }}
\def\ttwo{{\vartheta_2}}
\def\tthree{{\vartheta_3}}
\def\tfour{{\vartheta_4}}
\def\ti{{\vartheta_i}}
\def\tj{{\vartheta_j}}
\def\tk{{\vartheta_k}}
\def\calF{{\cal F}}
\def\smallmatrix#1#2#3#4{{ {{#1}~{#2}\choose{#3}~{#4}} }}
\def\ab{{\alpha\beta}}
\def\Minv{{ (M^{-1}_\ab)_{ij} }}
\def\bone{{\bf 1}}
\def\ii{{(i)}}
\def\V{{\bf V}}
\def\b{{\bf b}}
\def\N{{\bf N}}
\def\t#1#2{{ \Theta\left\lbrack \matrix{ {#1}\cr {#2}\cr }\right\rbrack }}
\def\C#1#2{{ C\left\lbrack \matrix{ {#1}\cr {#2}\cr }\right\rbrack }}
\def\tp#1#2{{ \Theta'\left\lbrack \matrix{ {#1}\cr {#2}\cr }\right\rbrack }}
\def\tpp#1#2{{ \Theta''\left\lbrack \matrix{ {#1}\cr {#2}\cr }\right\rbrack }}
\def\l{\langle}
\def\r{\rangle}

%================== BLACKBOARD BOLD CHARACTERS ==============================

\def\inbar{\,\vrule height1.5ex width.4pt depth0pt}

\def\IC{\relax\hbox{$\inbar\kern-.3em{\rm C}$}}
\def\IQ{\relax\hbox{$\inbar\kern-.3em{\rm Q}$}}
\def\IR{\relax{\rm I\kern-.18em R}}
 \font\cmss=cmss10 \font\cmsss=cmss10 at 7pt
\def\IZ{\relax\ifmmode\mathchoice
 {\hbox{\cmss Z\kern-.4em Z}}{\hbox{\cmss Z\kern-.4em Z}}
 {\lower.9pt\hbox{\cmsss Z\kern-.4em Z}}
 {\lower1.2pt\hbox{\cmsss Z\kern-.4em Z}}\else{\cmss Z\kern-.4em Z}\fi}

%========================================================================
%          MACROS FOR REFERENCES
%========================================================================
\def\AEF{A.E. Faraggi}
\def\NPB#1#2#3{{\it Nucl.\ Phys.}\/ {\bf B#1} (19#2) #3}
\def\PLB#1#2#3{{\it Phys.\ Lett.}\/ {\bf B#1} (19#2) #3}
\def\PRD#1#2#3{{\it Phys.\ Rev.}\/ {\bf D#1} (19#2) #3}
\def\PRL#1#2#3{{\it Phys.\ Rev.\ Lett.}\/ {\bf #1} (19#2) #3}
\def\PRT#1#2#3{{\it Phys.\ Rep.}\/ {\bf#1} (19#2) #3}
\def\MODA#1#2#3{{\it Mod.\ Phys.\ Lett.}\/ {\bf A#1} (19#2) #3}
\def\IJMP#1#2#3{{\it Int.\ J.\ Mod.\ Phys.}\/ {\bf A#1} (19#2) #3}
\def\nuvc#1#2#3{{\it Nuovo Cimento}\/ {\bf #1A} (#2) #3}
\def\etal{{\it et al\/}}

%==============================================================================
\hyphenation{su-per-sym-met-ric non-su-per-sym-met-ric}
\hyphenation{space-time-super-sym-met-ric}
\hyphenation{mod-u-lar mod-u-lar--in-var-i-ant}
%==============================================================================

%============================== SECTION 1 ============================

\setcounter{footnote}{0}
\section{Introduction}

Superstring theories \cite{Sreviews} are believed to provide a
consistent framework
for the unification of gravity with the gauge interactions.
An important task is to connect superstring theory with the
Standard Model \cite{heterotic,CHSW}.
Several approaches may be pursued to derive the Standard Model from
superstring theory. One possibility is to go through a simple
\cite{simple} or a semi--simple \cite{semisimple,revamp,ALR,FSU,LNY}
unifying gauge group
at intermediate energy scale. Another is to derive the Standard Model
directly from superstring theory \cite{ssm,fny,eu,SLM,custodial}.
Proton lifetime considerations motivate
the hypothesis that the Standard Model must be obtained directly from
superstring theory \cite{DTSMo,DTSMm}.
A second important question is
whether there exist some property of superstring models
that will distinguish them from other attempts to understand
the origin of the Standard Model. If such a property exists
it may result in an experimental signal that can prove or
disprove the validity of superstring unification.

In this paper we explore one such possible signature
of superstring unification.
We argue that realistic superstring models produce
additional heavy stable matter, beyond the spectrum of the
Standard Model.
The specific matter states and their properties vary between models.
However, the existence of additional stable matter, beyond
the observed spectrum of the Standard Model, is generic.
One type of such generic states in superstring models are
of course the moduli fields.
Indeed, it as been argued that because of the absence of
superpotential for the moduli fields,
they will decouple at a very early stage in the
evolution of the universe and will
overclose the universe \cite{modcos}. However, in the class of models
that we study, it has been
suggested that all the moduli  (except, of course, the dilaton)
are projected out by
the GSO projections \cite{halyo}. Thus, these models
the cosmological moduli problem can be resolved.
The matter states that we study in this paper arise due to
the superstringy breaking of the unifying gauge symmetry.
We investigate the possibility that these stringy stable matter states
can be the dark matter and can perhaps be detected.

In the attempts to derive the Standard Model from superstring theory
one traditionally starts with a larger, unifying, gauge symmetry $G$.
The gauge symmetry is then broken to the Standard Model by means
of Wilson lines. In many respects the unifying gauge symmetry $G$
is similar to the gauge group of four dimensional grand unification
and the Wilson lines are similar to the Higgs bosons in the
adjoint representation. However, there are some notable differences.
The eigenvalues of the Wilson lines are quantized while the
eigenvalues of the Higgs in the adjoint representation are continuous.
Another important difference is that the breaking of the gauge
symmetries by Wilson lines results in massless states that
do not fit into multiplets of the original unbroken
gauge symmetry. We refer to such states generically as
exotic ``Wilsonian'' matter states. This is an important
property as it may result in conserved quantum numbers that will
indicate the stability of these massless ``Wilsonian'' states.
The simplest example of this phenomenon is the existence
of states with fractional electric charge in the massless
spectrum of superstring models \cite{ww,eln,huet,fcp}.
Such states are stable due to electric charge conservation.
As there exist strong constraints on their masses and abundance,
states with fractional electric charge must be diluted away or
extremely massive.
Remarkably, however, the same ``Wilson line''
breaking mechanism, which produces matter with fractional electric
charge, is also responsible for the existence of states which
carry the ``standard'' charges under the Standard Model gauge
group but which carry fractional charges under a different subgroup
of the unifying gauge group. For example, if the group $G$ is
$SO(10)$ then the ``Wilsonian'' states carry non--standard
charges under the $U(1)_{Z^\prime}$ symmetry, which is embedded
in $SO(10)$ and is orthogonal to $U(1)_Y$.
Such states can therefore be stable if the $U(1)_{Z^\prime}$ gauge
symmetry remains unbroken down to low energies, or if some
residual local discrete symmetry is left unbroken after the
$U(1)_{Z^\prime}$ symmetry breaking.

In this paper we propose that the existence of heavy stable ``Wilsonian''
matter may be the ``smoking gun'' of string unification.
The existence of stable ``Wilsonian'' states at intermediate
energy scale have important cosmological implications. In a previous
letter \cite{letter} we examined the possibility that one type of
the extra ``Wilsonian'' states constitute the dark matter of the
universe. These states consist of heavy down--like quark with the
standard down--like charge assignment. Due to its role in the string
unification we referred to this type of particle as the {\it uniton}.
We proposed that because of its ``fractional'' charge under the
$U(1)_{Z^\prime}$ symmetry, the uniton may be stable.

In this paper we extend the analysis of ref. \cite{letter}.
We discuss in detail the cosmological constraints on the existence
of heavy ``Wilsonian'' states. We provide the details of the
analysis of ref. \cite{letter} and extend our investigation to
other exotic matter states which appear in the realistic superstring
derived models. In the superstring models that we consider the
unifying gauge symmetry is $SO(10)$. The $SO(10)$ symmetry is
broken at the string level to $SO(6)\times SO(4)$, $SU(5)\times U(1)$
or $SU(3)\times SU(2)\times U(1)^2$. We classify the exotic
``Wilsonian'' matter states according to the pattern of the
$SO(10)$ symmetry breaking. The $SO(6)\times SO(4)$ and
$SU(5)\times U(1)$ type models give rise to fractionally charged states
with $Q_{\rm e.m.}=\pm1/2$. On the other hand, the
$SU(3)\times SU(2)\times U(1)^2$ type models produce in addition
states with the regular charges under the Standard Model gauge group but
with ``fractional'' charges under the $U(1)_{Z^\prime}$ gauge group.
These states include down--like color triplets
and electroweak doublets, as well as states which are
Standard Model singlets. We show, by analyzing the renormalizable
and nonrenormalizable terms of the superpotential in a
specific superstring
model, that these exotic states can be stable.
We investigate the cosmological constraints on the masses and
relic density of the exotic states. We propose that, while
the abundance and the masses of the fractionally charged states
are highly constrained, the Standard Model -- like states,
and in particular the Standard Model singlet, are possible
candidates for the dark matter.

The exotic ``Wilsonian'' matter states that we study in this paper
are divided into three distinct classes:

The first class consists of down--like color triplets with ``fractional''
charge under the $U(1)_{Z^\prime}$ symmetry. The existence of such
a heavy colored state is motivated from the
constraints arising from string
gauge coupling unification \cite{DF}. In a specific superstring model we
analyze the interaction terms of this colored triplets with the
Standard Model states. In that model we show that if we assume that an
hidden $SU(3)_H$ gauge group remains unbroken then all the interaction
terms in the superpotential vanish to any order of
nonrenormalizable terms.
This result arises due to the fractional $U(1)_{Z^\prime}$ of the
color ``Wilsonian'' triplets and because in the specific string model
which we analyze in detail the only other Standard Model singlet
states which carry fractional $U(1)_{Z^\prime}$ charge are triplets of
$SU(3)_H$. Thus, we argue that the ``Wilsonian'' states can arise as
stable states. We then proceed to analyze the constraints on the
relic density of stable heavy color triplets. The heavy color triplets
can annihilate into quarks, squarks, gluons and gluinos.
We examine the possibility that the heavy color
down--like triplets are the dark matter. The heavy stable
down--like states form charged and neutral meson bound states
with the up and down quarks, respectively. An important issue
in this regard is the mass splitting between the charged and
neutral meson states. We argue that with our present understanding
of QCD, and the experimental determination of the light quark masses,
there exists a region in the parameter space in which the neutral
heavy meson state is the neutral one.

Next we examine the constraints on the relic density
of fractionally charged states, with electric charge $\pm1/2$.
We show that generically these states either have to be super massive
or have to be inflated away. We demonstrate in one specific model
that all the fractionally charged states have a cubic level mass term.
Thus, all the fractionally charged states can decouple from the massless
spectrum by some choices of flat directions. An alternative is that
all the fractionally charged states are confined by some non--Abelian
gauge group in the hidden sector. Another novel feature that arises
in some string models is the appearance of fractionally charged
baryons and fractionally charged leptons. Thus, one can speculate that
the these baryons and leptons will continue
to scatter in the early universe until they coalesce to
form neutral heavy hydrogen--like atoms.
%We briefly examine this scenario and conclude that .......

The final class of ``Wilsonian'' states that we consider are
Standard Model singlets with fractional $U(1)_{Z^\prime}$ charge.
These type of states arise in the superstring derived
standard--like models and interact with the Standard Model states
only via the $U(1)_{Z^\prime}$ gauge boson and are candidates for
weakly interacting dark matter (WIMPs). We examine four
possible scenarios: with and without inflation and with the
$U(1)_{Z^\prime}$ gauge boson being heavier or lighter than
the ``Wilsonian''--singlet states. We propose that this Standard Model
singlet states is the most likely candidate for the dark matter in the
superstring models.

Our paper is organized as follows. In section two we review the realistic
free fermionic superstring models. In section three we describe the
exotic ``Wilsonian'' states and classify them according to the patterns
of the $SO(10)$ symmetry breaking.
In section four we examine the cosmological
constraints on the different classes of ``Wilsonian''
matter states which are obtained from the superstring models.
In section five we present our conclusions. Our discussion of the
mass difference in the heavy meson system are given in appendix A.
The details of the calculation of the annihilation cross sections
are give in appendix B.

\setcounter{footnote}{0}
\section{Realistic free fermionic models}

In the free fermionic formulation of the heterotic string \cite{heterotic}
all the degrees of freedom needed to cancel the conformal anomaly
are represented in terms of internal free fermions propagating
on the string world--sheet.
In four dimensions, this requires 20 left--moving
and 44 right--moving real world--sheet fermions.
Equivalently, some real fermions may be paired to
form complex fermions.
Under parallel transport around a noncontractible loop,
the fermionic states pick up a phase.
Specification of the phases for all world--sheet fermions
around all noncontractible loops contributes
to the spin structure of the model.
Such spin structures are usually given in the form
of boundary condition ``vectors'', with each element
of the vector specifying the phase of a corresponding
world--sheet fermion.
The possible spin structures are constrained
by string consistency requirements
(e.g. modular invariance). A model is constructed by
choosing a set of boundary condition basis vectors,
which satisfies the modular invariance constraints.
The basis vectors, $b_k$, span a finite
additive group $\Xi=\sum_k{{n_k}{b_k}}$
where $n_k=0,\cdots,{{N_{z_k}}-1}$.
The physical massless states in the Hilbert space of a given sector
$\alpha\in{\Xi}$, are obtained by acting on the vacuum with
bosonic and fermionic operators and by
applying the generalized GSO projections. The $U(1)$
charges, $Q(f)$, with respect to the unbroken Cartan
generators of the four
dimensional gauge group, which are in
one--to--one correspondence with the $U(1)$
currents ${f^*}f$ for each complex fermion $f$, are given by:
\beq
{Q(f) = {1\over 2}\alpha(f) + F(f)}
\eeq
where $\alpha(f)$ is the boundary condition of the
world--sheet fermion $f$
 in the sector $\alpha$, and
$F_\alpha(f)$ is a fermion number operator counting each mode of
$f$ once (and if $f$ is complex, $f^*$ minus once).
For periodic fermions,
$\alpha(f)=1$, the vacuum is a spinor in order to represent the Clifford
algebra of the corresponding zero modes. For each
periodic complex fermion $f$
there are two degenerate vacua ${\vert +\rangle},{\vert -\rangle}$ ,
annihilated by the zero modes $f_0$ and
${{f_0}^*}$ and with fermion numbers  $F(f)=0,-1$, respectively.

The realistic models in the free fermionic formulation are generated by
a basis of boundary condition vectors for all world--sheet fermions
\cite{revamp,fny,ALR,eu,TOP,SLM,LNY,custodial}.
The basis is constructed in two stages. The first stage consists
of the NAHE set \cite{revamp,SLM},
which is a set of five boundary condition basis
vectors, $\{{{\bf 1},S,b_1,b_2,b_3}\}$. The gauge group after the NAHE set
is $SO(10)\times SO(6)^3\times E_8$ with $N=1$ space--time supersymmetry.
The vector $S$ is the supersymmetry generator and the superpartners of
the states from a given sector $\alpha$ are obtained from the sector
$S+\alpha$. The space--time vector bosons that generate the gauge group
arise from the Neveu--Schwarz sector and from the sector
$\xi=1+b_1+b_2+b_3$. The Neveu--Schwarz sector produces the generators of
$SO(10)\times SO(6)^3\times SO(16)$. The sector $\xi=1+b_1+b_2+b_3$
produces the spinorial 128 of $SO(16)$ and completes the hidden
gauge group to $E_8$.
The vectors $b_1$, $b_2$ and $b_3$ correspond to the three twisted
sectors in the corresponding orbifold formulation and produce
48 spinorial 16 of $SO(10)$, sixteen from each sector $b_1$,
$b_2$ and $b_3$.

The NAHE set divides the 44 right--moving and 20
left--moving real internal
fermions in the following way: ${\bar\psi}^{1,\cdots,5}$ are complex and
produce the observable $SO(10)$ symmetry; ${\bar\phi}^{1,\cdots,8}$ are
complex and produce the hidden $E_8$ gauge group;
$\{{\bar\eta}^1,{\bar y}^{3,\cdots,6}\}$, $\{{\bar\eta}^2,{\bar y}^{1,2}
,{\bar\omega}^{5,6}\}$, $\{{\bar\eta}^3,{\bar\omega}^{1,\cdots,4}\}$
 give rise to the three horizontal $SO(6)$ symmetries. The left--moving
$\{y,\omega\}$ states are divided to, $\{{y}^{3,\cdots,6}\}$, $\{{y}^{1,2}
,{\omega}^{5,6}\}$, $\{{\omega}^{1,\cdots,4}\}$. The left--moving
$\chi^{12},\chi^{34},\chi^{56}$ states carry the supersymmetry charges.
Each sector $b_1$, $b_2$ and $b_3$ carries periodic boundary conditions
under $(\psi^\mu\vert{\bar\psi}^{1,\cdots,5})$
and one of the three groups:
$(\chi_{12},\{y^{3,\cdots,6}\vert{\bar y}^{3,\cdots6}\},{\bar\eta}^1)$,
$(\chi_{34},\{y^{1,2},\omega^{5,6}\vert{\bar y}^{1,2}{\bar\omega}^{5,6}\},
{\bar\eta}^2)$ and $(\chi_{56},\{\omega^{1,\cdots,4}\vert{\bar\omega}^{1,
\cdots4}\},{\bar\eta}^3)$. The division of the internal fermions is a
reflection of the underlying $Z_2\times Z_2$ orbifold
compactification \cite{foc}. The set of internal fermions
${\{y,\omega\vert{\bar y},{\bar\omega}\}^{1,\cdots,6}}$
corresponds to the left--right symmetric conformal field theory of the
heterotic string, or to the six dimensional compactified manifold in a
bosonic formulation.

The second stage of the basis construction consists of adding three
additional basis vectors to the NAHE set. The allowed boundary
conditions in the additional basis vectors are constrained
by the string consistency constraints, i.e. modular invariance and
world--sheet supersymmetry. These three additional basis
vectors
correspond to ``Wilson lines'' in the orbifold formulation.
The additional basis vectors distinguish between different models
and determine their low energy properties.
Three additional vectors are needed to reduce the number of generations
to three, one from each sector $b_1$, $b_2$ and $b_3$.
At the same time the additional boundary condition basis vectors
break the gauge symmetries of the NAHE set.
The $SO(10)$ symmetry is broken to one of its
subgroups $SU(5)\times U(1)$, $SO(6)\times SO(4)$ or
$SU(3)\times SU(2)\times U(1)_{B-L}\times U(1)_{T_{3_R}}$.
This is achieved by the assignment of boundary conditions to the
set ${\bar\psi}^{1,\cdots,5}$:
\begin{eqnarray}
b\{{{\bar\psi}_{1\over2}^{1\cdots5}}\}=
\{{1\over2}{1\over2}{1\over2}{1\over2}
{1\over2}\}\Rightarrow SU(5)\times U(1),\label{so10to64}\\
b\{{{\bar\psi}_{1\over2}^{1\cdots5}}\}=\{1 1 1 0 0\}
  \Rightarrow SO(6)\times SO(4).\label{so10to51}
\end{eqnarray}
To break the $SO(10)$ symmetry to $SU(3)\times SU(2)\times
U(1)_C\times U(1)_L$\footnote{$U(1)_C={3\over2}U(1)_{B-L};
U(1)_L=2U(1)_{T_{3_R}}.$}
both steps, (\ref{so10to64}) and (\ref{so10to51}),
are used, in two separate basis vectors.
In the superstring derived standard--like models the three additional
basis vectors, beyond the NAHE set, are denoted by
$\{\alpha,\beta,\gamma\}$.
The two basis vectors $\alpha$ and $\beta$ break the $SO(10)$ symmetry
to $SO(6)\times SO(4)$ and the vector $\gamma$ breaks
the $SO(10)$ symmetry to $SU(5)\times U(1)$.
Since the standard--like superstring derived models contain
the $SO(6)\times SO(4)$ breaking sectors,
as well as the $SU(5)\times U(1)$
breaking sectors, their massless spectra admits also the exotic
representations that can appear in these models.
Therefore, below we will focus mostly on the superstring derived
standard--like models and comment on the overlap with the other models.

The observable gauge group after application
of the generalized GSO projections is
$SU(3)_C\times U(1)_C\times SU(2)_L\times U(1)_L
\times U(1)^3\times U(1)^n$.
The electromagnetic charge is given by
\begin{equation}
U(1)_{\rm e.m.}=T_{3_L}+U(1)_Y,
\label{quem}
\end{equation}
where $T_{3_L}$ is the diagonal
generator of $SU(2)_L$ and $U(1)_Y$ is the weak hypercharge.
The weak hypercharge is given by{\footnote{
Note that we could have instead defined the weak hypercharge to be
$U(1)_Y={1\over 3} U(1)_C - {1\over 2} U(1)_L$. This amounts to the same
redefinition of fields between the straight and flipped $SU(5)$. In this
paper we will use the definition in Eq. \ref{qu1y}.}}
\begin{equation}
U(1)_Y={1\over 3} U(1)_C + {1\over 2} U(1)_L
\label{qu1y}
\end{equation}
and the orthogonal
combination is given by
\begin{equation}
U(1)_{Z^\prime}= U(1)_C - U(1)_L.
\label{quzp}
\end{equation}
The horizontal $SO(6)^3$ symmetries are broken to factors of
$U(1)$s. The first three horizontal $U(1)$ symmetries arise from the
world--sheet complex fermions,
${\bar\eta}^j$, $(j=1,2,3)$. The additional $U(1)^n$ symmetries
arise from complexifying two right--moving real fermions from the set
$\{{\bar y},{\bar\omega}\}^{1,\cdots,6}$.
For each right--moving gauged $U(1)$ symmetry there is a corresponding
left--moving global $U(1)$ symmetry. Alternatively, a left--moving
real fermion can be paired with a right--moving real fermion to form
an Ising model operator \cite{KLN}. The hidden
gauge group after application of the generalized GSO projections is
$SU(5)_H\times SU(3)_H\times U(1)^2$.
The $U(1)$ symmetries in the
hidden sector, $U(1)_7$ and $U(1)_8$,
correspond to the world--sheet currents
${\bar\phi}^1{\bar\phi}^{1^*}-{\bar\phi}^8{\bar\phi}^{8^*}$ and
$-2{\bar\phi}^j{\bar\phi}^{j^*}+{\bar\phi}^1{\bar\phi}^{1^*}
+4{\bar\phi}^2{\bar\phi}^{2^*}+{\bar\phi}^8{\bar\phi}^{8^*}$ respectively,
where summation on $j=5,\cdots,7$ is implied.

The massless spectrum of the standard--like models contains
three chiral generations from the sectors $b_1$, $b_2$ and
$b_3$ with charges under the
horizontal symmetries. Three generations from the sectors
$b_1$, $b_2$ and $b_3$ are common to all the free fermionic
standard--like models. For example in the model of
Ref. \cite{gcu} we have,
\beqn
({e_L^c}+{u_L^c})_{{1\over2},0,0,{1\over2},0,0}+
({d_L^c}+{N_L^c})_{{1\over2},0,0,{{1\over2}},0,0}+
(L)_{{1\over2},0,0,-{1\over2},0,0}+(Q)_{{1\over2},
				0,0,-{1\over2},0,0},&&\nonumber\\
({e_L^c}+{u_L^c})_{0,{1\over2},0,0,{1\over2},0}+
({N_L^c}+{d_L^c})_{0,{1\over2},0,0,{1\over2},0}+
(L)_{0,{1\over2},0,0,-{1\over2},0}+
(Q)_{0,{1\over2},0,0,-{1\over2},0},&&\nonumber\\
({e_L^c}+{u_L^c})_{0,0,{1\over2},0,0,{1\over2}}+
({N_L^c}+{d_L^c})_{0,0,{1\over2},0,0,{1\over2}}+
(L)_{0,0,{1\over2},0,0,-{1\over2}}+
(Q)_{0,0,{1\over2},0,0,-{1\over2}}.&&\label{b123}
\eeqn
 with
\beqn
{e_L^c}&&\equiv ~[(1,{3\over2});(1,1)]_{(1,1/2,1)};\label{elc}\\
{u_L^c}&&\equiv ~[({\bar 3},-{1\over2});(1,-1)]_{(-2/3,1/2,-2/3)};
							\label{ulc}\\
Q&&\equiv ~[(3,{1\over2});(2,0)]_{(1/6,1/2,(2/3,-1/3))}\label{q}\\
{N_L^c}&&\equiv ~[(1,{3\over2});(1,-1)]_{(0,5/2,0)};\label{nlc}\\
{d_L^c}&&\equiv ~[({\bar 3},-{1\over2});(1,1)]_{(1/3,-3/2,1/3)};
							\label{dlc}\\
L&&\equiv ~[(1,-{3\over2});(2,0)]_{(-1/2,-3/2,(0,1))},\label{l}
\eeqn
where we have used the notation
\begin{equation}
[(SU(3)_C\times U(1)_C);
     (SU(2)_L\times U(1)_L)]_{(Q_Y,Q_{Z^\prime},Q_{\rm e.m.})},
\label{notation}
\end{equation}
and for the doublets we have written the electric charge of the two
components.
In the superstring--derived standard--like models the vectors
$b_1$, $b_2$ and $b_3$ are the only vectors in the additive group
$\Xi$ which give rise to spinorial $16$ of $SO(10)$.
In this case the three light generations are identified unambiguously
with the states from these three sectors.

The Neveu--Schwarz sector produces, in addition to the gravity
and the gauge boson multiplets, three pairs of electroweak doublets,
three pairs of $SO(10)$ singlets which are charged with respect to
$U(1)_{1,2,3}$, and three states which are singlets of the entire
four dimensional gauge group. In the realistic free fermionic models,
typically there is one additional sector which produces electroweak
doublet representations. Usually, this sector is a combination of
two of the vectors which extend the NAHE set.
For example, in the model ref. \cite{eu}, the combination
$b_1+b_2+\alpha+\beta$ produces
one pair of electroweak doublets, one pair of color triplets and five
pairs of $SO(10)$ singlets which are charged with respect to the $U(1)$
currents of the observable gauge group. All the states from the NS sector
as well as the states from the sector $b_1+b_2+\alpha+\beta$,
which carry Standard Model charges, are obtained by GSO projections from
the $10$ and $\overline{10}$ representation of $SO(10)$. Obviously,
the $SO(10)$ singlet fields from these sector do not carry any
Standard Model charges.

The states above complete the representations that we identify with
possible representations of the Standard Model.
In addition to the Standard
Model states, semi--realistic superstring models may contain additional
multiplets, in the $16$ and $\overline{16}$
representation of $SO(10)$, in
the vectorial $10$ representation of $SO(10)$,
or the $27$ and $\overline{27}$
of $E_6$. Such states can pair up to form super massive states.
They can mix with, and decay into, the Standard Model representation
unless some additional symmetry, which forbids their decay, is imposed.
For example, in the
flipped $SU(5)$ superstring models \cite{revamp},
two of the additional vectors
which extend the NAHE set produce an additional $16$ and $\overline{16}$
representation of $SO(10)$. These states are used in the flipped
$SU(5)$ model to break the $SU(5)\times U(1)$ symmetry to
$SU(3)\times SU(2)\times U(1)$.

All the states above fit into standard representation of the
grand unified group which may be, for example, $SO(10)$ or $E_6$.
They carry the standard charges under the Standard Model gauge
group or of its GUT extensions. The superstring models, however,
contain additional states that cannot fit into multiplets of the
original unifying gauge group. In the next section we enumerate the
states that appear in free fermionic models. These states
have important cosmological implications which we study below.

\setcounter{footnote}{0}
\section{ Exotic matter in free fermionic models}

At the level of the NAHE set the observable gauge symmetry is
$SO(10)\times SO(6)^3$ and the number of spinorial $16$ representations
of $SO(10)$ is 48. The basis vectors of the NAHE set are
seen to correspond to $Z_2\times Z_2$ orbifold compactification
at an enhanced symmetry point in the toroidal compactification
space \cite{foc}. To reduce the number of generations, and to break
the $SO(10)$ gauge
group to the Standard Model gauge symmetry, additional basis vectors
are added to the NAHE set. Three additional basis vectors are needed
to obtain three generations, one from each of the sectors $b_1$,
$b_2$ and $b_3$. The additional basis vectors which break the $SO(10)$
gauge symmetry correspond to Wilson lines in the orbifold formulation.

Adding to the untwisted sector the three twisted sectors $b_1$, $b_2$ and
$b_3$ results in additional massless states from these sectors.
In the same way adding the sectors $\{\alpha,\beta,\gamma\}$ to the NAHE
set results in additional massless spectrum from combinations
of these basis vectors with those of the NAHE set. However,
since these sectors correspond to ``Wilson lines'' they
give rise to massless states that do not fit into
representations of the original $SO(10)$ symmetry.
As a result the massless spectrum contains
states with fractional charges under the
unbroken $U(1)$ generators of the original
non--Abelian gauge group.
This is a new feature of superstring models.
Due to the absence of adjoint representations,
at least in superstring models with level one
gauge groups, ``Wilson line'' breaking is the only
available mechanism to break the unifying gauge
symmetry perturbatively.
Therefore, the appearance of massless states
with fractional $U(1)$ charges is a common phenomenon
in superstring models. In many examples the exotic
states appear in vector--like representations and can
acquire a heavy mass.
The ``Wilsonian'' matter phenomenon is an important feature
as it may result in discrete symmetries that may prevent the
decay of the exotic massive states into the Standard Model states.

The following exotic matter representations can appear in free
fermionic level one models. Sectors that break the
$SO(10)$ symmetry to $SO(6)\times SO(4)$ can contain the basis
vectors $\alpha$ or $\beta$. Sectors that break the
$SO(10)$ symmetry to $SU(5)\times U(1)$ contain the basis vector
$\gamma$ with a combination of the other basis vectors. We use the
following naming scheme for our particles. An exotic quark state will
be denoted by $W_{q_i}$, $W_{{\bar q}_i}$ while all the non--colored
exotic states will be denoted by $W_{\ell_i}$, $W_{{\ell}_i}$.

{}From the $SO(6)\times SO(4)$ type sectors we obtain the
following exotic states.

\bigskip

$\bullet$ {\underline {Color triplets:}}
\beqn
&&[(    3, {1\over2});(1,0)]_{( 1/6, 1/2, 1/6)}~~~~;\\
&&[({\bar3},-{1\over2});(1,0)]_{(-1/6,-1/2,-1/6)}
\label{sexton}
\eeqn
\parindent=15pt
\smallskip

Due to its fractional  charge under $U(1)_Y$
we refer to this state as the {\it sexton}. The sexton appears
for example in the model of ref. \cite{custodial} from the sector
$1+\alpha+2\gamma$ and in the model of ref. \cite{ALR} from the sector
$S+b_2+b_4+\alpha$. The sexton binds with light quarks to form
mesons and baryons with  fractional
electric charges $\pm1/2$ and $\pm3/2$.

\bigskip

$\bullet$ {\underline{Electroweak doublets:}}
\beq
[(1,0);(2,0)]_{(0,0,\pm1/2)}
\label{fc64doublet}
\eeq
\parindent=15pt

Such states appear for example in the model of ref.
\cite{custodial} from the sectors
$1+b_i+\alpha+2\gamma$ $(i=1,2,3)$.
In the model of ref. \cite{fny} from the sectors
$\xi+b_1+\alpha+2\gamma$ and $\xi+\alpha+2\gamma$,
and in the model of ref. \cite{ALR} from the sectors
$b_1+\alpha$, $b_1+b_2+b_4+\alpha$, $b_2+b_3+b_5+\alpha$,
$b_4+\alpha$, $b_1+b_4+b_5+\alpha$.

We also obtain from this type of sectors fractionally charged
$SU(3)_C\times SU(2)_L$ singlets. In the $SO(6)\times SO(4)$ models
these states are doublets of $SU(2)_R$ which have zero $U(1)_C$ charge
and the $SU(3)_C$ singlet in the quartets of $SU(4)$ with zero $U(1)_L$
charge. In the standard--like models these states are
$SU(3)_C\times SU(2)_L$ singlets with electric charge
$Q_{\rm e.m.}=\pm1/2$.

\bigskip

$\bullet$ {\underline{Fractionally charged
		$SU(3)_C\times SU(2)_L$ singlets :}}
\beqn
&&[(1,0);(1,\pm{1})]_{(\pm1/2,\mp1/2,\pm1/2)}\\
&&[(1,\pm3/2);(1,0)]_{(\pm1/2,\pm1/2,\pm1/2)}
\label{fc64singlet}
\eeqn
\parindent=15pt

{}From sectors which break the $SO(10)$ symmetry into $SU(5)\times U(1)$
we obtain exotic states with fractional electric charge $\pm1/2$

\bigskip

$\bullet$ {\underline{Fractionally charged $SU(3)_C\times SU(2)_L$ singlets :}}
\beq
[(1,\pm3/4);(1,\pm{1/2})]_{(\pm1/2,\pm1/4,\pm1/2)}
\label{fc51singlet}
\eeq
\parindent=15pt

In general the fractionally charged states may transform
under a non--Abelian hidden gauge group in which case the fractionally
charged states may be confined.
For example, in the ``revamped'' flipped $SU(5)$ model \cite{revamp}
the states with fractional charge $\pm1/2$ transform as $4$ and $\bar4$
of the hidden $SU(4)$ gauge group. In other models these states may
be singlets of all the non--Abelian group factors.
Such states appear for example in the model of ref.
\cite{eu} from the sectors
$\{b_1+b_2+\alpha+\beta\pm\gamma,
   b_1+b_3+\alpha+\beta\pm\gamma,
   b_2+b_3+\alpha+\beta\pm\gamma,
   b_1+b_2+b_3+\alpha+\beta\pm\gamma\}$.
In the model of ref. \cite{fny}, they appear from the sector
$\{\pm\gamma,
    \xi\pm\gamma,
    1+b_4\pm\gamma,
    \xi+1+b_4\pm\gamma,
    \xi+b_3\alpha\pm\gamma\}$.

Finally in the superstring derived standard--like models
we may obtain exotic states from sectors which are combinations
of the $SO(6)\times SO(4)$ breaking vectors and $SU(5)\times U(1)$
breaking vectors. These states then carry the standard charges
under the Standard Model gauge group but carry fractional charges
under the $U(1)_{Z^\prime}$ gauge group.
The following exotic states are obtained:

\bigskip

$\bullet$ {\underline{color triplets :}}

\beqn
&&[(3,{1\over4});(1,{1\over2})]_{(-1/3,-1/4,-1/3)}; \\
&&[(\bar3,-{1\over4});(1,{1\over2})]_{(1/3,1/4,1/3)}
\label{uniton}
\eeqn
\parindent=15pt

In ref. \cite{letter}, due to its potential role in string gauge
coupling unification \cite{gcu,DF},
we referred to this state as ``the uniton''.
Such states appear for example in the model of ref. \cite{eu}
from the sector $b_2+b_3+\alpha\pm\gamma$,
in the model of ref.
\cite{custodial} from the sectors $ b_{1,2}+b_3+\beta\pm\gamma$,
and in the model of ref. \cite{fny} from the sectors
$\{b_3+\alpha\pm\gamma,
        b_1+b_2+b_4+\alpha\pm\gamma\}$.

\bigskip

$\bullet$ {\underline{electroweak doublets : }}

\beq
[(1,\pm{3\over4});(2,\pm{1\over2})]_{(\pm1/2,\pm1/4,(1,0);(0,-1))} \\
\label{fc321doublet}
\eeq

\parindent=15pt

Unlike the previous electroweak doublets, these electroweak doublets
carry the regular charges under the standard model gauge group but carry
``fractional'' charge under the $SO(10)$ symmetry. Electroweak
doublets of these type appear for example in the model of ref.
\cite{eu} from the sector $b_1+b_3+\alpha\pm\gamma$.
In the model of ref. \cite{fny} from the sectors
$\{b_3+\alpha\pm\gamma,
        b_1+b_2+b_4+\alpha\pm\gamma\}$.

Finally, in the superstring derived standard--like models we also obtain
states which are Standard Model singlets and carry ``fractional''
charges under the $U(1)_{Z^\prime}$ symmetry.

\bigskip

$\bullet$ {\underline{Standard model singlets with ``fractional''
			$U(1)_{Z^\prime}$ charge :}}

\beq
[(1,\pm{3\over4});(1,\mp{1\over2})]_{(0,\pm5/4,0)}
\label{fc321singlet}
\eeq

\parindent=15pt

These states may transform under a non--Abelian hidden gauge group
or may be singlets of all the non--Abelian group factors.
This type of Standard Model singlet appears in all the
known free fermionic standard--like models. For example, 0
in the model of ref.
\cite{custodial} they are obtained from the sectors
$b_{1,2}+b_3+\beta\pm\gamma$.
In the model of ref. \cite{fny} they appear from the sectors
$\{b_1+b_2+b_3+\alpha\pm\gamma,
   b_2+b_3+b_4+\alpha\pm\gamma,
   1+b_1+b_2+\alpha\pm\gamma,
   1+b_3+b_4+\alpha\pm\gamma,
   b_3+\alpha\pm\gamma,
   b_1+b_2+b_4+\alpha\pm\beta\}$.

There are several important issues that are important to examine
with regard to the exotic states. Since some of these
states carry fractional charges, it is desirable to make them
sufficiently heavy or sufficiently rare.
All the exotic matter states appear in vector--like representations.
They can therefore obtain mass terms from renormalizable or higher order terms
in the superpotential. We must then study the renormalizable and
nonrenormalizable superpotential in the specific models.
The cubic level and higher order terms in the superpotential are
extracted by applying the rules of ref. \cite{KLN}.
The problem of fractionally charged states
was investigated in ref. \cite{fcp} for the model of ref.
\cite{fny}. In the notation of ref. \cite{fny}
the massless states $\{V_{41},V_{42},V_{43},V_{44},
		       V_{47},V_{48},V_{49},V_{50}\}$ are of
			the form of Eq. (\ref{fc64singlet});
the massless states $\{V_{45},V_{46},V_{51},V_{52}\}$ are of the form of
			Eq. (\ref{fc64doublet});
the massless states $\{H_1-H_{14}\}$ are of the form
			of Eq. (\ref{fc51singlet});
the massless colored states $\{H_{33},H_{40}\}$ are of the form
			of Eq. (\ref{uniton}) and the
the massless weak doublets $\{H_{34},H_{41}\}$ are of the form
			of Eq. (\ref{fc321doublet}).
The remaining exotic states, $\{H_{15}-H_{32},H_{35}-H_{39},H_{42}\}$,
are Standard Model singlets of the form
			of Eq. (\ref{fc321singlet}).
The cubic level superpotential of the exotic massless states is given by
\begin{eqnarray}
W_2&=&{1\over{\sqrt2}}\{H_1H_2\phi_4+H_3H_4{\bar\phi}_4+
H_5H_6{\bar\phi}_4+(H_7H_8+H_9H_{10})\phi_4'\nonumber\\
&+&(H_{11}+H_{12})(H_{13}+H_{14}){\bar\phi}_4'
 +V_{41}V_{42}{\bar\phi}_4+V_{43}V_{44}{\bar\phi}_4\nonumber\\
&+&V_{45}V_{46}\phi_4+(V_{47}V_{48}+V_{49}V_{50}){\bar\phi}_4'
+V_{51}V_{52}\phi_4'\}\nonumber\\
&+&[H_{15}H_{16}\phi_{56}'
+H_{17}H_{18}{\bar\phi}_{56}+H_{19}H_{20}{\bar\phi}_{56}'+
H_{21}H_{22}{\bar\phi}_{56}\nonumber\\
&+&(V_{11}V_{12}+V_{13}V_{14})\phi_{13}+
(V_{15}+V_{16})(V_{17}+V_{18})\phi_{13}+V_{19}V_{20}\phi_{13}\nonumber\\
&+&V_{21}V_{22}\phi_{12}+V_{23}V_{24}\phi_{12}+
(V_{25}+V_{26})(V_{27}+V_{28})\phi_{12}+V_{29}V_{30}\phi_{12}\nonumber\\
&+&V_{31}V_{32}{\bar\phi}_{23}
+V_{33}V_{34}\phi_{23}+H_{29}H_{30}{\bar\phi}_{13}+
H_{36}H_{37}\phi_{12}].\label{fnysuperp}
\end{eqnarray}
By examining the fractionally charged states and the trilinear
superpotential, it is observed that all the fractionally charged
states receive a Planck scale mass by giving a VEV to
the neutral singlets
${\bar\phi}_4,{\bar\phi}_4',{\phi}_4,\phi_4'$
which imposes  the additional F flatness constraint
$(\phi_4{\bar\phi}_4'+{\bar\phi}_4\phi_4')=0$. The other exotic states
which are Standard Model singlets do not receive mass by this choice
of flat direction. Therefore, at this level of the superpotential, the
fractionally charged states can decouple from the remaining light
spectrum. Similarly, the issue of fractionally charged states in the
model of ref. \cite{eu} was studied in ref. \cite{nrt} where it was
found that all the fractionally charged states receive large mass from
renormalizable or nonrenormalizable terms. Similar results
were also found in the case of the Gepner models \cite{huet}.

The second issue that must be examined with regard to the
exotic ``Wilsonian'' matter is the interactions with the Standard Model
states. The fractional charges of the exotic states
under the unbroken $U(1)$ generators of the $SO(10)$ gauge group,
may result in conserved discrete symmetry which forbid their
decay to the lighter Standard Model states. In the following
we will investigate this question with regard to particular states
that appear in specific models.

\newcommand{\beqa}{\begin{eqnarray}}
\newcommand{\eeqa}{\end{eqnarray}}
\newcommand{\sq}{\tilde{q}}
\newcommand{\sg}{\tilde{g}}
\newcommand{\sbb}{\tilde{b}}
\newcommand{\dnot}[1]{\not{\!\! #1}}

\section{Cosmology of ``Wilsonian'' matter}
\subsection{Introduction}
The ``Wilsonian'' matter states obtained from
the realistic superstring models are heavy and stable.
Their mass density has important cosmological implications,
since if they are too abundant they will overclose the universe.
This gives a limit on their present relic density $\rho_0$
\begin{equation}
\rho_0 \leq \rho_c = 1.054 h^2\times 10^4\mbox{eV cm}^{-3},
\end{equation}
where $h$ is defined by the present value of the Hubble constant $H_0$
\begin{equation}
h=\frac{H_0}{100 \rm ~km ~sec^{-1} ~Mpc^{-1}}.
\end{equation}

The relic density of a particle depends on its mass, couplings
and on the reheating temperature after inflation.
If the relic density is the same as (or at least comparable to) the
critical density of the universe, it could account for
the existence of dark matter in the universe.

Since the ``Wilsonian'' matter states are heavy,
they are non--relativistic at the time of structure formation and are
candidates for {\em cold} dark matter.
Models consisting of only hot dark matter, such as neutrinos of a few eV
mass, are ruled out phenomenologically,
since they fail to explain the large--scale structure curve
of the universe.
Only models which contain cold dark matter, like
the mixed (cold $+$ hot)
dark matter model \cite{MDM} and the cold dark matter
with extra radiation model \cite{DWR}, survive, except
for models with non--standard cosmology scenarios,
such as models with non--zero cosmological constant.
The heavy ``Wilsonian'' matter states are candidates for
cold dark matter. The possible existence of heavy stable particles
in the realistic superstring models provides further motivation to
study this class of models.

Since these ``Wilsonian'' states cannot decay into normal particles,
their number density can only change by annihilation processes.
In several examples that we study below, the ``Wilsonian'' states
are interacting strongly and therefore remain in thermal equilibrium
until they become non--relativistic. We will consider one exception of
a $SU(3)\times SU(2)\times U(1)_Y$ singlet which interacts weakly.
A given ``Wilsonian'' matter state decouples from the thermal bath when
its annihilation rate falls below the expansion rate of the universe.
The annihilation rate of a particle is given by
\begin{eqnarray}
\Gamma =
 \langle\sigma_{\rm ann}|v|\rangle n_{_{EQ}},
\end{eqnarray}
where $n_{_{EQ}}$, the number density at the equilibrium,
is given by
\begin{equation}
{\displaystyle{n_{_{EQ}}}}=\cases{
{\displaystyle{g_{\rm eff}\left(\frac{\zeta(3)}{\pi^2}\right)T^3}}&
			$\hbox{~~~relativistic}$\cr
{\displaystyle{g_{\rm eff}\left(\frac{mT}{2\pi}\right)^{3/2}\exp(-M/T)}}&
	      		$\hbox{~~~non-relativistic}$\cr}
\end{equation}
Here $\zeta(3)=1.20206$ is the Riemann zeta function of 3,
and $g_{\rm eff}$ is the effective number of
degrees of freedom of the particle.

In the expanding universe, the evolution equation of the
particle number density is described by the Boltzmann equation
\begin{equation}
\frac{dn}{dt}+3Hn=-\langle \sigma_{\rm ann} |v|\rangle (n^2-n_{_{EQ}}^2),
\label{boltz}
\end{equation}
where the Hubble constant in the radiation dominated era
is a function of the temperature
\begin{eqnarray}
H =  1.66 \sqrt{g_*}\frac{T^2}{m_{pl}}.
\end{eqnarray}
During the radiation dominated era,
time ($t$) and temperature are related by
\begin{equation}
t=0.301 g^{-1/2}_* \frac{m_{pl}}{T^2}.
\end{equation}
It is standard to define the dimensionless parameters $x=M/T$
and the number density in a comoving volume
$Y=n/{\rm s}_e$, where ${\rm s}_e$ is the entropy density
\begin{equation}
{\rm s}_e =\frac{2\pi^2}{45} g_{*s} M^3 x^{-3}.
\end{equation}

Here $g_*$ and $g_{*s}$ are the relativistic degrees of
freedom and are defined by
\begin{eqnarray}
g_* &=  &\sum_{i=bosons}g_i\left(\frac{T_i}{T}\right)^4 +
\frac{7}{8}\sum_{i=fermions}g_i\left(\frac{T_i}{T}\right)^4,\nonumber  \\
g_{*s} &=  &\sum_{i=bosons}g_i\left(\frac{T_i}{T}\right)^3 +
\frac{7}{8}\sum_{i=fermions}g_i\left(\frac{T_i}{T}\right)^3,
\end{eqnarray}
where $g_i$ and $T_i$ are the degrees of freedom and the
temperature of a particle $i$, respectively.
The Boltzmann equation, Eq. (\ref{boltz}), describes the
evolution of the number density with respect to temperature
in a comoving volume,
\begin{equation}
\frac{dY}{dx} = -\lambda x^{-2} (Y^2-Y^2_{EQ}),
\label{boltz2}
\end{equation}
where
\begin{eqnarray}
\lambda=\frac{x\langle\sigma_{\rm ann}|v|\rangle {\rm s}_e}{H}.
\end{eqnarray}
In many cases $\lambda$, in the non--relativistic limit,
becomes $x$-independent. Equation (\ref{boltz2}) can be solved if
we know the annihilation cross section as a function of the temperature.
It is known that the
relic density of a particle of mass $M$ is given by two
different expressions, depending upon the two possible
temperature regimes at which decoupling occurs
\begin{equation}
Y_0 = \cases{
{\displaystyle{0.278\left(\frac{g_{\rm eff}}
			{g_{*s(T_{ dec})}}\right)}}&$~~~M\ll T_{dec}$\cr
{}&${}$\cr
{\displaystyle{\frac{3.79 x_{dec}}{\sqrt{g_*} m_{pl}
			M\langle\sigma|v|\rangle}}}&$~~~M>T_{dec}$\cr}
\label{rel}
\end{equation}
where we have defined $x_{dec}=M/T_{dec}$, and
$T_{dec}$ denotes the decoupling temperature.
The two cases $M\ll T_{dec}$ and $M>T_{dec}$ describe
decoupling in relativistic and non--relativistic
regimes, respectively. If we define s$_{e_0}$ to be the entropy
density of the present universe, then
the relic energy density of a massive decoupled
particle is given by
\begin{equation}
\rho_0={\rm s}_{e_0} Y_0 M = 2.97\times 10^3 Y_0 M~\mbox{cm}^{-3},
\end{equation}
and we can estimate the $\Omega_0$ parameter,
which is given by the expression

\begin{equation}
\Omega_0 h^2 \equiv \frac{\rho_0 h^2}{\rho_c}=\frac{2970 ~M ~Y_0 ~{\rm
cm}^{-3}}
{1.05 \times 10^4 ~{\rm eV ~cm}^{-3}}\,\,.
\label{mass}
\end{equation}
Using the fact that the cosmological data set the bound
$0.1 <\Omega_{0} h^2<1 $,
Eq.~(\ref{mass}) gives a bound on the mass of a stable particle
\begin{equation}
M < \frac{3.5}{Y_0} {\rm eV}. \label{bound}
\end{equation}

If a massive particle is weakly interacting (WIMP)
at high temperature, it
can decouple from thermal equilibrium when it is still relativistic.
In this case the upper bound for its mass is very low, since the $Y_0$
value is too high. This can be deduced from Eq.~(\ref{rel}). In the case
of an inflationary scenario, however, this bound can be raised. We will
elaborate over this possibility in some detail in the following sections.

The class of superstring models which we have discussed in the previous
sections allow several kinds of stable ``Wilsonian''
matter states. In the
following we will specifically focus our attention
on three cosmologically
interesting cases.

\subsection{A stable heavy down--like quark: the uniton}

In this section we investigate the possibility that the
dark matter is composed of heavy down--like color triplets
of the form of Eq.~(\ref{uniton}).
This will be our most detailed quantitative and complex
analysis, and the discussion
with regard to the other potential dark matter candidates from
superstring models will be somewhat more qualitative.

\setcounter{footnote}{0}
\subsubsection{Motivation: string gauge coupling unification}

The existence of additional colored vector--like states beyond the spectrum
of the Minimal Supersymmetric Standard Model is motivated also from
a different consideration.
Superstring unification predicts that the gauge and gravitational
couplings are unified at the string
unification scale which is of the order
$M_{\rm string}\approx g_{\rm string}5\cdot 10^{17}~{\rm GeV}$
with $g_{\rm string}\sim0.8$ \cite{Ginsparg,scales,Kaplunovsky}.
Assuming, naively, that the massless states
below the string scale consists solely of the
Minimal Supersymmetric Standard Model results
in disagreement with the experimentally observed values
for $\sin^2\theta_W(M_Z)$ and $\alpha_{\rm strong}(M_Z)$.
If one assumes naively that the spectrum between the electroweak scale
and the unification scale consists solely of the MSSM state
then the gauge couplings are seen to intersect at
$\sim2\times10^{16}~{\rm GeV}$ \cite{gcumssm}.
This discrepancy is usually referred to as the order of magnitude
mismatch between the MSSM and string unification scales.

It would seem that in an extrapolation of the gauge parameters
over fifteen orders of magnitude, a problem involving a single order
of magnitude would have many possible resolutions. Indeed, in string theory
there are many possible effects that can {\it a priori} account for the
discrepancy. For example, there could be large corrections to the gauge
couplings due to the infinite tower of heavy string modes \cite{moduli}.
Alternatively, additional matter \cite{Gaillard,gcu,LNY,DF} or gauge
structure beyond the MSSM could reconcile the two scales. Yet another
possibility is that the weak hypercharge normalization in string
theory is not necessarily the one that is traditionally found in GUTs
and could take values that would yield meeting of the couplings at
the string scale \cite{ibanez}.

Surprisingly, however, the discrepancy is not easily resolved.
In ref. \cite{DF} the string--scale gauge coupling unification
problem was investigated in detail in the context of the realistic free
fermionic superstring models. It was shown,
in a wide range of realistic free
fermionic models, that heavy string threshold corrections, non--standard
hypercharge normalizations,
light SUSY thresholds or intermediate gauge
structure do not resolve the problem. Instead, the problem may only be
resolved due to the existence of additional
intermediate matter thresholds,
beyond the MSSM spectrum. This additional matter takes the form of
additional color triplets and electroweak doublets, in vector--like
representations. Remarkably, some string models contain in their massless
spectrum the additional states with the specific weak hypercharge
assignments, needed to achieve string
scale unification \cite{gcu}. Perhaps
even more intriguing is the fact that in some models the additional
states are those that arise due to the ``Wilson line'' breaking.
Thus, the existence of stable ``Wilsonian'' states at intermediate
energy scales is motivated also from requiring consistency of the
string--scale gauge coupling unification with the observed
low energy gauge parameters.

The mass scale of the additional color triplets is constrained
by requiring agreement between the low energy gauge observables and
unification of the gauge couplings at the string scale.
In ref. \cite{DF}
the following two constraint equations on the
intermediate matter mass scales
were obtained,
\beqn
   & 10.29 ~<~ \sum_i \,(b_{2_i}-b_{1_i})\,
\ln {\displaystyle {M_S\over{M_i}}}
    ~<~ 14.14~& \label{imte1}\\
   &  18.57~<~\sum_i\,(b_{3_i}-b_{1_i})\,
			\ln{\displaystyle {M_S\over{M_i}}}~<~ 30.58~,
\label{imte2}
\eeqn
where $M_S$ is the string unification scale, $M_i$ are the mass scales
of the intermediate thresholds, and $\{b_3,b_2,b_1\}_i$
are the corresponding
one--loop $\beta$--function coefficients of
$SU(3)_C\times SU(2)_L\times U(1)_Y$. Eqs. (\ref{imte1}), (\ref{imte2})
can now be used to constrain the mass scale of the additional color
triplets. A variety of possible combinations exist. For
example, in the model of ref. \cite{gcu}
with one pair of light color triplets of the uniton type,
$\{D_1,\overline{D}_1\}$, we obtain the limits
\beq
      {\rm experimental~limit}~<~M_3~<~ 1.81~{\rm TeV}~.
\label{abcdefg}
\eeq
Setting $M_3$ at the upper limit of (\ref{abcdefg}),
we find
\beq
     7.2\,\times\,10^{13}~{\rm GeV}~<~ M_2 ~<~2.6\,\times\,10^{14}~{\rm GeV}~,
\eeq
while for a lower limit of $M_3\sim500$ GeV we obtain
\beq
            3.6\,\times\,10^{5}~{\rm GeV}~<~M_2~<~1.7\,\times\,10^{6}~{\rm
GeV}~.
\eeq
By contrast, with {\it two}\/ triplet pairs of the uniton type,
$\{D_1,\overline{D}_1,D_2,\overline{D}_2\}$,
degenerate at one mass scale $M_3$,
we
instead find
\beq
                4.3\,\times\,10^{6}~{\rm GeV}~<~ M_3 ~<~ 9.5\,\times\,
10^{10}~{\rm GeV}~,
\eeq
so that taking the upper limit for $M_3$ yields
\beq
           7.2\,\times\, 10^{13}~{\rm GeV}~<~ M_2 ~<~ 2.6\,\times\,
10^{14}~{\rm GeV}~
\eeq
while the lower limit on $M_3$ yields
\beq
               5\,\times\, 10^{12}~{\rm GeV}~<~ M_2
        ~<~1.8\,\times\,10^{13}~{\rm GeV}~.
\eeq
Finally, adding the third pair of color triplets of the sexton type
$\{D_3,\overline{D}_3\}$, and with all three color triplet pairs
degenerate at the scale $M_3$, we find
\beq
               2.4\,\times\,10^{11}~{\rm GeV}~<~ M_3~<~
7.2\,\times\,10^{13}~{\rm GeV}
\eeq
for which the upper and lower limits respectively yield
\beqn
    3.7\,\times\, 10^{14}~{\rm GeV}~&<~ (M_2)_{\rm upper}~&<~ 1.1\,\times\,
10^{15}~{\rm GeV}~\nonumber\\
    5.7\,\times\, 10^{13}~{\rm GeV}~&<~ (M_2)_{\rm lower}~&<~ 2\,\times\,
10^{14}~{\rm GeV} ~.
\eeqn
Clearly, many viable scenarios exist, and the above examples are
not exhaustive.

Next we observe that these mass scales can be compatible with the
flat directions of the cubic level superpotential. For example, in the model
of ref. \cite{gcu}, the relevant cubic level mass terms are
\begin{eqnarray}
&+&{1\over2}(\xi_1D_1{\bar D}_1+\xi_2D_2{\bar D}_2)\nonumber\\
&+&{1\over\sqrt{2}}(D_1{\bar D}_2\phi_2+{\bar D}_1D_2{\bar\phi}_1)\}
\label{d1d2}
\end{eqnarray}
and there exist $F$ and $D$ flat vacua in which the VEV of the singlets
in (Eq. \ref{d1d2}) vanishes. We therefore conclude that indeed there
is sufficient freedom in the superstring models that allows the additional
color triplets to appear at intermediate mass scales.

\setcounter{footnote}{0}
\subsubsection{Interactions}

We now discuss the interactions of the uniton with the
Standard Model states. In the model of ref. \cite{eu} there is
one pair of uniton states denoted by $\{H_{21},H_{22}\}$ and one
pair of exotic doublets denoted by $\{H_{15},H_{16}\}$.
The following cubic level superpotential terms are obtained
\begin{equation}
{D_{45}}{H_{18}}{H_{21}}+h_2H_{16}H_{17}+
{h_{45}}{H_{16}}{H_{25}}+
			{1\over2}(\xi_1H_{21}H_{22}
                        +\xi_2H_{15}H_{16}).
\label{eu1sp}
\end{equation}

The states $\{{H_{18}},H_{17},{H_{25}}\}$ are Standard Model singlets
with ``fractional'' $U(1)_{Z^\prime}$ charge of
the form of Eq. (\ref{fc321singlet}).
{}From Eq. (\ref{eu1sp}) we observe that in the model of ref. \cite{eu},
there are no interactions terms, at order $N=3$,
with the states from the sectors $b_1$, $b_2$ and $b_3$.
However, potential interaction terms do appear
for the exotic color triplets
and electroweak doublets with the color triplets from the
sector $b_1+b_2+\alpha+\beta$ and the electroweak doublets from the
Neveu--Schwarz sector. These interaction terms are generated by the
breaking of $U(1)_{Z^\prime}$. For specific choices of flat directions
these terms vanish. However,
potential non--vanishing interaction terms may be
generated from nonrenormalizable terms.
The uniton would be a more appealing
dark matter candidate if there existed a gauge symmetry or a local
discrete symmetry \cite{KW} that forbids the interactions terms
to all orders of nonrenormalizable terms.

Next we turn to the model of ref. \cite{custodial}. The superpotential
terms of the exotic color triplets with the Standard Model states
were studied in ref. \cite{DTSMm}.
The potential interaction terms are
\begin{eqnarray}
&&LQ{\bar D},~u_L^ce_L^cD,~QQD,~u_L^cd_L^c{\bar D},~d_L^cN_L^cD,\label{sm2u1}\\
&&QDh\label{sm1u1}\\
&&{\bar D}{\bar D}u_L^c\label{sm1u2}
\end{eqnarray}

For the terms in Eq. (\ref{sm2u1}) the type of
correlators that have to be checked are of the form
$b_ib_j{D}\phi^n$, where $b_i$ and $b_j$ represent states from
the sectors $b_i$ and $b_j$, $D$ are the additional color triplets,
and $\phi^n$ is a string of Standard Model singlets.
For the first two pairs of color triplets from the sectors
$b_{1,2}+b_3+\alpha+\beta$,
the operators $b_ib_j{D}$ are invariant under the weak hypercharge.
However, they break $U(1)_{Z^\prime}$ because
$Q_{Z^\prime}({D})={(1/2)}Q_{Z^\prime}({D_{45}})$.
Thus, $D$ has one half the $U(1)_{Z^\prime}$ charge of the triplets
from the Neveu--Schwarz and $b_1+b_2+\alpha+\beta$ sectors.
Therefore, all the operators in Eq. (\ref{sm2u1}), with $D$
being a triplet from one of
the sectors $b_{1,2}+b_3+\alpha+\beta$, break $U(1)_{Z^\prime}$.
Thus, the string $\l\phi\r^n$
contains a $U(1)_{Z^\prime}$ breaking VEV. However, in this model
all the available Standard Model singlets with nontrivial
$U(1)_{Z^\prime}$ charge transform as $3$ and ${\bar 3}$ of
the Hidden $SU(3)$ gauge group \cite{gcu}.
The $U(1)_{Z^\prime}$ charges of the hidden $SU(3)$ triplets are
$\pm5/4$. The $U(1)_{Z^\prime}$ charges of the color triplets from the
exotic ``Wilson line'' sectors are $\pm1/4$ (see table 1).
The last pair of color triplets has ``fractional'' weak hypercharge
$Q_Y=\pm1/12$.
Consequently, terms of the form of Eq. (\ref{sm2u1}),
with $D$ being a triplet from one of
the exotic ``Wilson line'' sectors, cannot be formed in this model.
This result was verified by a computer search of the relevant
nonrenormalizable terms up to order $N=12$.

The term in Eq. (\ref{sm1u1}) breaks $U(1)_{Z^\prime}$.
Therefore, the product $\phi^n$ in a potential
non--vanishing higher order term must break $U(1)_{Z^\prime}$.
The available fields are
$\{H_1,H_2,{\bar H}_1,{\bar H}_2\}$. These fields transform
as $3$ and $\bar3$ of the hidden $SU(3)_H$
gauge group with opposite charges under $U(1)_{Z^\prime}$
for the states and the bared states.
Consequently, to obtain terms which are invariant under
the hidden $SU(3)_H$  gauge group,
some of the $\{V_{1,2,3}, {\bar V}_{1,2,3}\}$ fields,
which transform as $3$ and ${\bar3}$
of $SU(3)_H$,  must get a VEV.  Therefore if we impose
that the VEVs of the $V_i$,  ${\bar V}_i$
are suppressed, the terms of the of Eq. (\ref{sm1u1})
are suppressed to all orders of
nonrenormalizable terms. We remark that there is no phenomenological
constraint that requires the VEVs of $V_i$,  ${\bar V}_i$
to be non--vanishing.
In contrast to the model of ref. \cite{eu} in
which generation mixing is obtained
by these VEVs \cite{CKM}, in this model the
generation mixing is obtained
by the VEVs of the  fields $T_i$,  ${\bar T}_i$.
Finally, the term in Eq. (\ref{sm1u2}) is invariant under $U(1)_Y$
and $U(1)_{Z^\prime}$. However the product
${\bar D}{\bar D}$ carries nontrivial
charges under $U(1)_7$ and $U(1)_8$. All the Standard
Model singlet fields which
are charged under $U(1)_7$ and $U(1)_8$ transform
as $5$, ${\bar 5}$ of $SU(5)_H$ or as
$3$, ${\bar3}$ of $SU(3)_H$. Consequently, the product
$\phi^n$ must contain a baryonic
factor under $SU(5)_H$ or $SU(3)_H$. Such terms were
not found up to order
$N=12$. Imposing that the VEVs of $V_i$,
${\bar V}_i$  are suppressed, suppresses the
terms of the form of Eq. (\ref{sm1u2}), because
the $U(1)_{7,8}$ charges of the product
$5_H\cdots5_H$ cannot be canceled. Thus,
with this assumption the term
${\bar D}{\bar D} u_L^c$ is forbidden to all orders
of nonrenormalizable terms. To summarize, if we assume that
$SU(3)_H$ remains unbroken at the Planck scale, it is seen
that all the interaction terms in Eqs. (\ref{sm2u1},\ref{sm1u1},\ref{sm1u2})
vanish, to all orders or nonrenormalizable terms.

\setcounter{footnote}{0}
\subsubsection{Uniton dark matter}

In the previous sections we have seen that superstring models
in the free fermionic formulation motivate the existence of
vector--like heavy stable $SU(3)_C$ triplets, $(3,1)_{1/3}$,
$(3,1)_{1/6}$. In this notation the second quantum number
denotes their $SU(2)_L$ content and the subscript denotes
the hypercharge. The uniton \cite{letter}, i.e. the state
$(3,1)_{1/3}$, is a $d$--type quark and interacts with the
color force. In the mass range we are interested in
($10^3<M< 10^{14}$ GeV), the interaction rate is greater than
the expansion rate of the universe, unless the number density of these
particles is significantly suppressed. In general it is well
known that it is hard to envision a scenario in which
a particle carrying unbroken gauge charges decouples
in a relativistic regime.
Thus, the uniton can be decoupled
only when it becomes non--relativistic.
In the non--relativistic limit $T/M<1$, the
annihilation rate of the uniton is given by
\begin{eqnarray}
\Gamma \simeq\frac{\pi N \alpha_s^2 }{M^2 } n_{_{EQ}},
\end{eqnarray}
where $M$ is the mass of the uniton and
$\alpha_s\equiv g_s^2/4\pi$ is the gauge
coupling at the decoupling temperature. In this equation
$n_{_{EQ}}$ is the number density of the uniton in
the non--relativistic limit, while
$N$ is obtained by summing over all the
available annihilation channels and is given by
\begin{equation}
N=\sum_f{c_f}.
\end{equation}
A discussion of this result can be found in appendix B where we present
a complete calculation of all the main channels of annihilation of this
particle. The amplitudes $c_f$ are obtained
by taking the non--relativistic
limit of the cross sections. The calculation is implemented
in the case of $N=1$ supersymmetric QCD with heavy quark-antiquark
($Q\bar{Q}$) initial states and light gluinos ($\tilde{g}$), squarks
($\tilde{q}$), quarks ($q$) and gluons ($g$) in the final
state ($Q\bar{Q}\to q\bar{q},\,
Q\bar{Q}\to g\,g,\,Q\bar{Q}\to \tilde{q}\tilde{q},\,
Q\bar{Q}\to \tilde{g}\tilde{g}$).
We refer to figs. 5-10 for details.

We obtain $c_q=4/3$ for quarks, $c_g=14/27$ for gluons,
$c_{\tilde q}=2/3$ for squarks and $c_{\tilde g}=64/27$ for gluinos.
Using these results we get the expression for
$\lambda$ (see Eq.~(\ref{boltz2}))
\begin{equation}
\lambda =0.83 N\alpha_s^2\frac{g_*s}{\sqrt{g_*}}\frac{m_{pl}}{ M} .
\end{equation}
At this point we can impose the decoupling condition
$dY/dx \simeq 0$ \cite{KTB} which gives
\begin{equation}
x_{dec} = \ln [(2+c)\lambda ac] -\frac{1}{2}\ln\{\ln [(2+c)\lambda ac]\},
\end{equation}
where $ a= 0.145(g/g_{*s})$ and $c$ is
$Y(T_{dec})/Y_{EQ}(T_{dec})$, the latter being
of order one. Therefore we estimate a decoupling temperature of the form
\begin{equation}
T_{dec}\simeq\frac{M}{\ln(m_{pl}/M)}.
\end{equation}

Using these results we can estimate the value of the
number density in a comoving volume in the present universe
\begin{equation}
Y_0~=~\frac{3.79 x_{dec}}{\sqrt{g_*} m_{pl} M \langle \sigma|v|\rangle}
   ~=~ 3.8\frac{M\ln(m_{pl}/M)}{N\alpha_s^2\sqrt{g_*}m_{pl}}.
\label{noinf1}
\end{equation}
We have set $g_* =g_{*s}$, since the decoupling temperature is high
\cite{KTB}.

Using this condition and Eq.~(\ref{bound}),
we get an upper bound on the mass of the uniton
\begin{equation}
M <10^5 \alpha_s \left(N\sqrt{g_*}\ln(m_{pl}/M)\right)^{1/2}\mbox{GeV}.
\label{noinf}
\end{equation}

We now turn to discuss the case in which we have inflation.
The effect of inflation is important if the decoupling
temperature is greater than the reheating temperature ($T_{dec} \gg T_R$).
If $T_R\ll T_{dec}$ the uniton will be diluted away,
although it can be regenerated after reheating by
{\em out-of-equilibrium} production \cite{KTB}.
Therefore we can approximately set the initial density of the uniton to
be zero and obtain the relation
\begin{equation}
\frac{dY}{dx} =  \lambda x^{-2}Y_{EQ}^2 \label{boltzmann2},
\end{equation}
with $Y_{EQ}= 0.145 g_{eff}/g_* x^{3/2} e^{-x}$.
Integrating this equation from the reheating
temperature down to the present temperature we get
\begin{equation}
Y_0= \frac{\lambda g_{eff}^2}{2}
 \left(\frac{0.145}{g_*}\right)^2 \left(x_r+\frac{1}{2}\right) e^{-2x_r},
\label{infla2}
\end{equation}
where $x_r\equiv M/T_R$ and
\begin{equation}
\Omega_0h^2 \simeq 9\times 10^3 N\alpha_s^2 g_{eff}^2\frac{m_{pl}}{\rm eV}
\left(\frac{200}{g_*}\right)^{1.5} \left(x_r+\frac{1}{2}\right) e^{-2x_r},
\end{equation}
from which we derive a bound on the mass of the uniton
\begin{equation}
M> T_R\left[25+\frac{1}{2}\ln\left(\frac{M}{T_R}\right)\right].
\label{yinf}
\end{equation}
Without inflation, we have  a strict bound on the mass of the uniton
which is around $10^5$ GeV.
However, inflation can raise the mass bound to any arbitrary order,
depending upon the estimated value of the reheating temperature.

In ref.~\cite{NR90} it was pointed out that an exotic stable
heavy quark, which decouples at high temperature, can
annihilate after color confinement.
At low temperature heavy quarks will form bound states of
finite size $a_f$ with ordinary quarks.
Since the temperature at confinement is quite small compared to
the heavy quark mass, the scattering cross section of two of these
heavy hadrons  will approximately be equal to their geometric cross section
$\sigma\sim 4\pi a_f^2$, with $a_f$  about 1 fm. We cannot estimate the
corresponding annihilation cross section at this temperature.
However, it is not unreasonable to assume that the two cross sections
(annihilation and scattering) are proportional to each other:
\begin{equation}
\sigma_{\rm ann} = \eta 4\pi a_f^2.
\end{equation}
Within this assumption, from the Boltzmann equation
we can calculate the value of the relic density
\begin{eqnarray}
\frac{dY}{dx}&=&-\frac{\langle \sigma |v| \rangle s}{xH}Y^2
\nonumber\\
&=& - M m_{pl} \eta \pi a_f^2 x^{-2.5}Y^2\,\,.
\label{conf1}
\end{eqnarray}
Notice that we have set $Y_{EQ}$ to be zero since
the temperature is much lower than the mass of the uniton.

After integration from confinement temperature to the present temperature,
we get
\begin{equation}
\frac{1}{Y_f}-\frac{1}{Y_i} = \frac{2}{3}M m_{pl} \eta \pi a_f^2
\left[\left(\frac{T_i}{M}\right)^{3/2}-\left(\frac{T_f}{M}\right)^{3/2}
\right]
\label{conf}
\end{equation}
with $i$ and $f$ characterizing the beginning and end of the
annihilation period, respectively.

If we assume that this mechanism can give a large suppression to
the relic density of the uniton, then we can ignore $T_f/M$ and $1/Y_i$
in Eq.~(\ref{conf}) and the new relic density becomes
\begin{equation}
Y_0\simeq \frac{3\sqrt{M}}{2\pi\eta a_f^2 m_{pl} T_i^{3/2}}\,\,.
\label{reequi}
\end{equation}
At this point we need an estimate of $\eta$.
If we assume that the scattering and the annihilation cross sections
are of same order, i.e. $\eta\simeq 1$,  and if we insert the value
$T_i\sim 1$ GeV, which is the temperature at the confining phase,
the upper bound on the uniton mass can be raised up to $10^7-10^8$ GeV.
However, we do not have any definitive argument which can help us
 estimate the value of $\eta$. Therefore this issue remains open.

If we ignore this re-equilibrium process,
then the mass of the uniton should be
smaller than $10^5$ GeV or greater than
 ${\cal O}(10)\times$ reheating temperature, which is the result presented
in Eqs.~(\ref{noinf}) and (\ref{yinf}).

On the other hand, the uniton can make a stable neutral bound state
and as such can be component of dark matter.
This issue is crucially connected to the question of
whether the neutral ($U_0$) or the the charged ($U_{-1}$)
bound state is the lightest between the two possible states of this meson.
We refer to appendix A for a discussion of this point. There we use
arguments based both on QCD potential models and results
from the heavy quark effective theory to conclude that the possibility of
having $U_0$ as the lowest state is not ruled out. The splitting between the
two states is likely to be of $1-2$ MeV, which is of the order
of the electromagnetic mass splitting,
and allows a conversion of $U_{-1}$ into
$U_0$ by beta decay.

If $U_0$ is a component of dark matter then its relic energy density
should have the same order of magnitude as the critical energy
density $\rho_c$. Therefore from Eq.~(\ref{mass})
we can estimate the mass of the $U_0$ both in the case of inflation and
without.
In the first case we get
\begin{equation}
M \simeq 10^5 \alpha_s \left(N\sqrt{g_*}\ln(m_{pl}/M)\right)^{1/2}\mbox{GeV},
\end{equation}
while in the latter case the estimated mass is given by
\begin{equation}
M\simeq T_R\left[25+\frac{1}{2}\ln\left(\frac{M}{T_R}\right)\right].
\end{equation}

We remark that there are three windows (in the parameter space
$M/\sigma_p$, with $\sigma_p$ denoting the scattering cross section
on protons)
for strongly interacting dark
matter (such as $U_0$) \cite{SGED90}
which possibly meet our requirements.
The first window is
in the relatively low mass range  (10 GeV $< M < 10^4$ GeV )
and in the range $10^{-24}<\sigma_p< 10^{-20}$cm$^2$.
In other two windows it is required to have $10^5$ GeV $< M <10^7$ GeV
and $M>10^{10}$ GeV, respectively,
assuming a cross section, in both cases, less than
$10^{-25} $cm$^2$.
These constraints include bounds from various experiments
(such as experiments performed using solid state cosmic-ray
detectors and plastic track cosmic-ray detectors)
and from cosmological consideration
(such as the galactic halo infall rate and the life-time
of neutron stars \cite{neutron}).

\subsection{Fractionally charged matter}
As we have discussed in the previous sections, among the ``Wilsonian''
states predicted by the free fermionic superstring models in the low
energy limit, there are also heavy particles which are
$SU(2)_L$ doublets and singlets
(and which are $SU(3)$ singlets),
$(1,2)_0$, $(1,1)_{1/2}$.
These two states are lepton--like, are stable and
have fractional electric charge $\pm 1/2$.
The behavior of these two states
at high temperature is supposed to be the same as in the uniton case since
the three gauge couplings become of the same order.

We have also seen that superstring models contain
fractionally charged vector--like quarks $(3,1)_{1/6}$.
We refer to this particle as the sexton.
This particle is expected to form bound states with $u,d$ quarks (since it is
a color triplet) and can make stable baryons and mesons with fractional
electric charge of $\pm 1/2$ and $\pm 3/2$.
The experimental searches for free quarks in various materials show
that the upper bound on the number density of fractionally charged particle
should be smaller than $10^{-19}\sim 10^{-26}$\cite{data}.
This implies that the relic density of fractionally charged matter
can be estimated to be
\begin{equation}
Y^{FC}_0 < 10^{-19} Y^B_0 < 10^{-19}\frac{{\rm eV}}{m_p}
 \sim 10^{-28}.
\label{so}
\end{equation}
This result almost excludes the possibility of fractionally
charged dark matter
since from Eq.~(\ref{bound}), the lower bound of the
masses of fractionally charged
states should be given by
\begin{equation}
M\sim \frac{{\rm eV}}{Y_0^{FC}}> 10^{19} {\rm GeV}.
\end{equation}
Since the sexton is a color triplet, we can estimate its relic
density $Y_0$ by the same way as in the uniton case (see (\ref{noinf1}) and
(\ref{infla2}).
It is clear that the relic density of the sexton  cannot satisfy
the constraint given in Eq.~(\ref{so}) without inflation if
the mass of the sexton is greater than 1 eV.
Instead, if the mass is lighter than 10 eV, then the sexton behaves like
a light quark ($u, d$) and is confined in a hadron bound state.
However, the densities of fractionally charged
hadron bound state at low temperature are severely constrained \cite{data}.
Therefore, string models which predict light stable fractionally
charged particles cannot survive without inflation,
since all these fractionally charged particles should be diluted away.
Another possible scenario is that the states which carry fractional charges
also transform under a non--Abelian gauge group in the hidden sector
in which case the fractionally charged states are confined \cite{revamp}.
In this case the fractionally charged states form bound states which are
integrally charged \cite{eln}. It should be noted, however, that in this
scenario one must insure that the neutron star constraints \cite{neutron}
are not violated.

{}From Eq.~(\ref{infla2})
we can also derive a bound on the relic density
of these particles in an inflationary scenario,
\begin{equation}
Y_0=\frac{\lambda g^2_{eff}}{2}
 \left(\frac{0.145}{g_*}\right)^2 \left(x_r+\frac{1}{2}\right) e^{-2x_r}<
10^{-28}.
\end{equation}
In this equation $\lambda$ is approximately the same as for the
uniton case $\sim N \alpha^2 \sqrt{g_*} m_{pl}/M$.
Therefore we have the approximate bound,
\begin{equation}
\frac{M}{T_R} > 38 + \log\left(\frac{T_R}{10^9 \rm GeV}\right).
\end{equation}

In order to evade the experimental constraints on the relic densities of
fractionally charged bound states, it is mandatory to show that the density of
these states is suppressed. Our arguments on this matter are quite
different from those presented in the earlier literature \cite{NR90}.
Two new elements appear in our analysis:
$a)$ the value of the charge of the sexton ($\pm 1/6$),
which is different from the usual quark charge assignment
($\pm 1/3, \pm 2/3$);
and $b)$ the presence of fractionally charged leptons.

At the confinement temperature, the sexton (denoted ${\sigma}$)
can form neutral and charged color singlet bound states
($\sigma\sigma\sigma, q\,\sigma\,\sigma,
 q\, q\, \sigma ,
\bar{q}\,\sigma$) and their corresponding antiparticle
($\bar{\sigma}\,\bar{\sigma}\,\bar{\sigma},
\bar{q}\,\bar{\sigma}\,\bar{\sigma},
\bar{q}\,\bar{q}\,\bar{\sigma}, q \,\bar{\sigma}$).
It is not hard to show that the number of fractionally charged bound states
which can be generated  in hadronic reactions is not small.
For instance we can classify all the possible conversion processes
for both integer and fractional final states.
Typical examples of these processes are
\beqa
&& q\,q\sigma + q\,q\, \sigma\,\, ({\rm frac.})\to q\,\sigma\,\sigma +
q\,q\,q\,\, ({\rm int.})
 \nonumber \\
&& \bar{q}\,\sigma + q\,q\, \sigma \,\,({\rm frac.})
\to q \sigma\,\sigma + \bar{q} \,q\, \,({\rm int.})\nonumber \\
&& \bar{q}\,\sigma + q\, \bar{\sigma}\, \,({\rm frac.})\to
\sigma\, \bar{\sigma} + q\,\bar{q}\,\,({\rm int.})
\nonumber \\
&& q\, q\, \sigma + q\,\bar{\sigma}\, \,({\rm frac.})\to
\sigma\, \bar{\sigma} + q\,q\,q\,
\,({\rm int.})\nonumber \\
\label{conv}
\eeqa
and their converses.
Integer charged final states can be easily reconverted into
fractionally charged states due to the large amount of ordinary particles
($q\,\bar{q}$ and $q\,q\,q\,$) present in the thermal bath.
As we can see from (\ref{conv}), annihilation channels
$(\sigma \, \bar{\sigma})$ are always present, which in turn
can reduce the total number
of sexton bound states. This number has been estimated in the case of the
uniton in the previous sections. As the temperature goes down due to  the
expansion of the universe, it is possible that the remaining fractional
hadrons will form bound states of integer charge with fractionally charged
heavy leptons $(1,2)_0$ and $(1,1)_{1/2}$
(in our conventions $Q_{\rm e.m.}= T_3 +Y$), therefore making neutral
heavy hydrogen--type bound states (e.g. $B_{1/2} + L_{-1/2}\to H_0$).
We can estimate the cross section for this process as follows.

At a temperature of few eV, the fractionally charged bound state $B_{1/2}$
or $L_{-1/2}$ can
capture an electron and form a bound state of radius $a_B$ (Bohr radius)
which is hydrogen--like fractionally charged bound state ($B_{1/2} + e^{-}$).
Therefore
we can assume that the cross section for two of these fractionally
charged bound state to interact and form a neutral bound state
is proportional to the scattering cross section of two hydrogen-like atoms
\begin{equation}
\sigma = \eta 4\pi a_B^2.
\end{equation}
Then we can calculate the new relic density from the Boltzmann
Eq.~(\ref{boltz2}), assuming that $Y_{EQ}$ is very small due to the
exponential suppression of their number density
\begin{eqnarray}
\frac{dY}{dx}= - M m_{pl} \eta \pi a_B^2 x^{-2.5}Y^2.
\end{eqnarray}
Notice that this the same as Eq.~(\ref{conf1}), except that $a_f$ is now
replaced
by $a_B$. Therefore we obtain
\begin{equation}
\frac{1}{Y_f}-\frac{1}{Y_i} = \frac{2}{3}M m_{pl} \eta \pi a_B^2
\left[\left(\frac{T_i}{M}\right)^{3/2}-\left(\frac{T_f}{M}\right)^{3/2}
\right]\,.
\label{cu}
\end{equation}

This result is the same as in Eq.~(\ref{conf}), but
with a lower temperature value $T_i$
and a larger geometrical size ($a_B > \,a_f$).
$Y_f$ here denotes the final density of fractionally charged particles
whose value is strongly restricted by experimental data \cite{data}.
If  this  neutral bound state $H_0$ is a dark matter candidate, it
should have the maximally allowed value of $Y_i$,
i.e.~the critical density is
\begin{equation}
Y_i\simeq \frac{{\rm 3.5eV}}{M},
\end{equation}
and if we insert this value into (\ref{cu}) we obtain the conversion rate
of fractionally charged bound states to the neutral ones
\begin{equation}
\frac{Y_i}{Y_f} -1 \sim \eta \left( \frac{10^6{\rm GeV}}{M}\frac{T_i}{10 {\rm
eV}}
\right)^{1.5}.
\label{soso}
\end{equation}
If we assume, in the
most optimistic scenario,
that $\eta\approx 1$, from Eqs.~(\ref{so}) and (\ref{soso}) we easily deduce
that a large $Y_i/Y_f$ ratio is allowed only if the mass $M$ of this bound
state is quite small ($10^{-20}$ GeV).
This is clearly unrealistic since
our calculation is valid only if these particles are non--relativistic.
Therefore we conclude that light fractionally charged
particles cannot be dark matter candidates.

{}From Eq.~(\ref{cu})
more generally, we can deduce the relic density of the fractionally
charged particles in the form
\begin{equation}
Y_f\simeq \frac{3\sqrt{M}}{2\pi\eta a_B^2 m_{pl} T_i^{3/2}}\,\,.
\label{reequi2}
\end{equation}
Here, we ignore $1/Y_i$ compared with $1/Y_f$, assuming that this
scattering process can suppress the relic density.
If we insert the values $a_B\simeq 2.7\times 10^5$ GeV$^{-1}$ and $T_i
\simeq 10$ eV, it cannot satisfy the constraint (\ref{so})
unless its mass is much smaller than the temperature $T_i\simeq 10$ eV.
Since a light particle with non--zero charge has not been detected
in the laboratory experiments \cite{data}, it is hard to imagine that
there is a strong suppression on the relic density of the fractionally
charged matter at low temperature.
These arguments seem to indicate that string models which contain light
fractionally charged states always need inflation. Finally, we comment
that fractionally charged states with Planck scale mass could perhaps
constitute the dark matter, but in this case the tools that we have
used for the analysis may not be adequate.

\subsection{A singlet of $SU(3)_c\times SU(2)_L \times U(1)_Y$}

Another ``Wilsonian'' state which is obtained in the superstring models,
and which is a possible candidate for dark matter
is a singlet of $SU(3)_c\times SU(2)_L \times U(1)_Y$,
carrying an additional non--standard $U(1)_{Z^\prime}$ charge.
This type of ``Wilsonian'' matter state arises due to the breaking
of the $SO(10)$ symmetry to $SU(3)\times SU(2)\times U(1)^2$
and appears generically in the superstring derived standard--like models.
For example, in the model of refs. \cite{gcu,custodial} such states
appear from the sectors $b_{1,2}+b_3+\beta\pm\gamma$.
In this model these states transform as $3$ and ${\bar 3}$ of
a hidden $SU(3)_H$ gauge group. As in section 4.2.2 their
interactions with the Standard Model states vanish to all
orders of nonrenormalizable terms if the $SU(3)_H$ is left unbroken.

The $U(1)_{Z^\prime}$ symmetry should be broken somewhere in
between the weak scale and the
Planck scale. In this section we refer to this state as the
``W--singlet''($W_s$).

After symmetry breaking the W--singlet interaction is
suppressed by $1/M^2_{Z^\prime}$ and can be classified as a weakly
interacting massive particle (WIMP).

The W--singlet can annihilate into two light Standard Model fermions
and into their superpartners. The latter can decay afterwards
into two $Z^\prime$ gauge bosons.
This decay will be suppressed if the gauge boson mass is greater than
the mass of W--singlet and this singlet, therefore, will be a stable particle.

Let $M$ be the W--singlet mass, and $M_{Z^\prime}$ the mass of the $Z^\prime$
gauge boson. Let us consider the total cross section for the
annihilation of two  W--singlet into two
normal fermions and their superpartners.

\begin{figure}
\centerline{\epsfbox{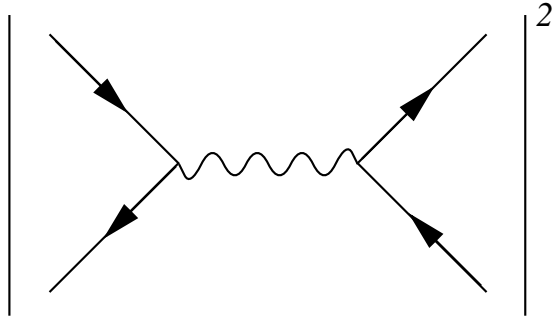}}\label{Ws2f}
\caption{$W_s\bar{W}_s\longrightarrow f\bar{f}$ decay}
\end{figure}

\begin{figure}
\centerline{\epsfbox{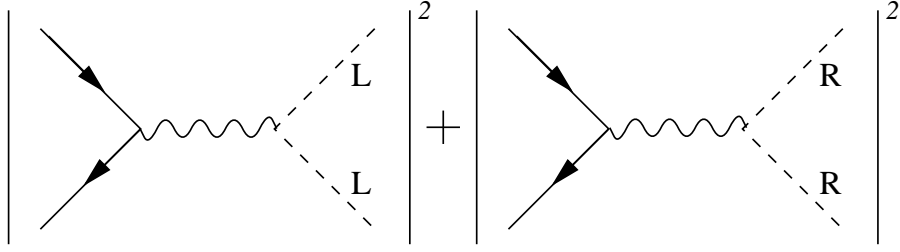}}\label{Ws2sf}
\caption{$W_s\bar{W}_s\longrightarrow \tilde{f}\tilde{f}^*$ decay}
\end{figure}

The two diagrams in figs.~1 and 2
are similar to those
shown in figs.~9 and 10, except for the color factors and for the mass of the
s--channel gauge boson, which is now assumed to have a mass $M_{Z^\prime}$.
We get
\begin{equation}
\sigma=\frac{4\pi N_{Z^\prime} \sqrt{{s}} ({s}+2M^2)}
{3 (M_{Z^\prime}^2-{s})^2
\sqrt{{s}-4M^2}}\, ,
\end{equation}
where $s$ is the square of the C.M. energy ${s}=4E_{CM}^2$ of each incoming
particle and
\begin{equation}
N_{Z^\prime}= \frac{Q^2_w}{(4\pi)^2}\sum_f Q^2_f.
\end{equation}
In this equation
$Q_i$ are the $U(1)_{Z^\prime}$
charges of the W--singlet and of the Standard Model
particle $f$. Notice that we have to introduce a
$1/2$ suppression factor for each scalar partner if
there is no mixing between the left--handed sfermion
and the right--handed sfermion.

Let us comment on the calculation of this cross section, which can be obtained
with a slight modifications of the results of Appendix B.
For this purpose we start distinguishing two cases: that of a heavy $Z'$
and that of a light $Z'$.
In the case of a heavy $Z'$ the only relevant diagrams are
those of figs.~1 and 2.
Notice that we do not allow, in this case, $Z'$'s in the final state.
Notice also that the annihilation of $W_s$ into regular quarks and leptons
(and into their supersymmetric partners) is allowed,
since all the quarks carry an extra $U(1)_{Z'}$ charge.
In the case of a leptophobic $Z'$ charge \cite{leptophobic1,leptophobic2}
we would include in the final states only color states.
\begin{figure}
\centerline{\epsfbox{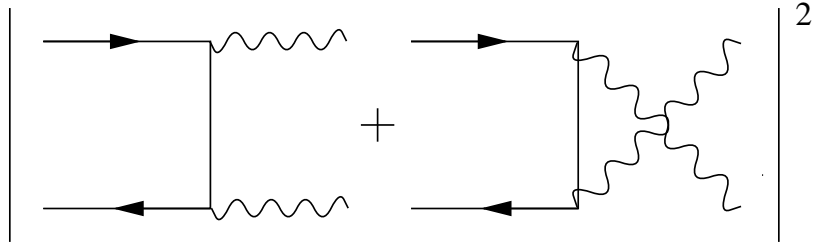}}\label{Ws2z}
\caption{$W_s\bar{W}_s\longrightarrow Z' Z'$ decay}
\end{figure}

\begin{figure}
\centerline{\epsfbox{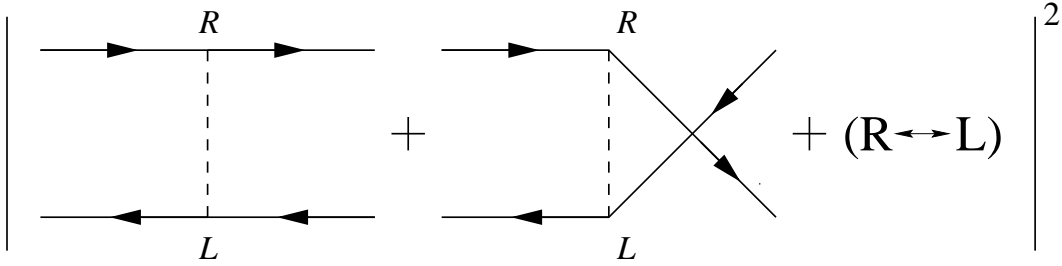}}\label{Ws2sz}
\caption{$W_s\bar{W}_s\longrightarrow \tilde{Z}'\tilde{Z}'$ decay}
\end{figure}

In the second case, when $Z'$ is light, we include also the channel
$W_s\,W_s \to 2 \,Z'$ beside those considered before.
The cross section for this last channel is shown in fig.~3
and 4.

The annihilation cross section into two $Z^\prime$ is given by
\begin{equation}
\sigma=\frac{4\pi\alpha_w^2}{{s} A^2}\left[ ({s}+4M^2)A -(
s^2+4M^2{s}-8M^4)
\ln\left(\frac{{s}+A}{{s}-A}\right)\right],
\end{equation}
and the one into two gauginos
\begin{equation}
\sigma=\frac{4\pi\alpha_w^2}{A^2}\left[ A + 2M^2
\ln\left(\frac{{s}+A}{{s}-A}\right)\right],
\end{equation}
where $A=\sqrt{{s}({s}-4M^2)}$.

Since the decoupling temperature of WIMP's
depends on both the mass and coupling
constant, we cannot use the same estimate
as in the uniton case.

Here we consider four different cases.

\bigskip
1) \underline{W--singlet mass greater than $M_{Z^\prime}$ without inflation.}

In this case, the W--singlet is a strongly interacting particle when it
decouples from the heat bath and therefore the relic density estimate is
the same as in the uniton case.
The $W_s$ is non--relativistic at decoupling and its total
annihilation cross section
into two $Z^\prime$s and into two gauginos is given by
\begin{equation}
\sigma|v| = (1+2)\frac{2\pi}{M^2} \alpha_w^2,
\end{equation}
while the corresponding two--fermion/sfermion cross section is
\begin{equation}
\sigma|v| = \frac{\pi}{M^2} N_{Z^\prime},
\end{equation}
where $\alpha_w=Q_w^2/4\pi$.

Although the $W_s$'s are
strongly interacting particles at decoupling from
the thermal bath, they
interact weakly in the present universe since the interaction
of this particle with the
Standard Model particles goes only through the $U(1)_{Z'}$
gauge bosons. This interaction, at the present epoch, is suppressed
by the heavy $Z'$ gauge boson mass since the symmetry is broken.
Therefore the bounds on the three windows for
the strongly interacting dark matter which we discussed above
(section 4.2) are not valid in this case.
This leads to constraints which are similar to those of the uniton
\begin{equation}
M <10^5 \left((N_z+2\alpha_w^2)\sqrt{g_*}\ln(m_{pl}/M)\right)^{1/2}\mbox{GeV}.
\end{equation}

\bigskip
2) \underline{W--singlet mass greater than $M_{Z^\prime}$ with inflation.}

If the reheating temperature is greater than $M_{Z^\prime}$,
the estimated cross section is the same as for the first case.
Using Eq.~(\ref{infla2}) similarly
we can estimate the relic density of the $W_s$
in an inflationary scenario and obtain a bound on its mass
\begin{equation}
M> T_R\left[25+\frac{1}{2}\ln\left(\frac{M}{T_R}\right)\right].
\end{equation}
However, if the reheating temperature is less than
$M_{Z^\prime}$, the regeneration of $W_s$ goes through a weak $U(1)_{Z'}$
process and the bound given above is not valid.
This scenario will be discussed in the fourth case below.

\bigskip
3) \underline{W--singlet mass is less than $M_{Z^\prime}$ and no inflation.}

In this case the $W_s$ is a WIMP. Therefore we will assume that
decoupling will occur when it
is still relativistic.
Their number density in the comoving volume is then
estimated to be
\begin{equation}
Y_0\equiv \frac{n_{_{EQ}}}{{\rm s}}=
 0.278\frac{g_{\rm eff}}{g_{*s(T_{\rm dec})}} \simeq 1.2\times10^{-3}.
\label{ws3}
\end{equation}
In Eq. (\ref{ws3}) we assumed that the
particle content is that of the MSSM and that
the decoupling temperature is larger than $1~{\rm TeV}$. This is a
conservative assumption as it will minimize the value of $Y_0$ and
hence maximize the mass bound.
Using this value of $Y_0$ in Eq.~(\ref{mass})
we can derive the upper mass bound
\begin{equation}
M< 3 {\rm keV}.
\end{equation}
The $W_s$ with a few keV can be a candidate of warm dark matter.

\bigskip
4) \underline{W--singlet mass less than $M_{Z^\prime}$ with inflation.}

A heavy mass WIMP can be a dark matter candidate only in the presence of
inflation. In fact, inflation will dilute away all the existing W--singlets
which will then be regenerated after reheating.
In this limit, i.e. $M<M_{Z^\prime}$ and $T_R<M_{Z^\prime}$, we
approximate the $Z^\prime$--mediated interaction by a four--point
Fermi interaction. In this temperature regime,
a $W_s$ ${\bar{W}}_s$ pair cannot
annihilate into two $Z^\prime$s because it is too massive.
We also observe
that after supersymmetry breaking the mass of the superpartners of the
$Z^\prime$s is of the same order (modulo soft breaking corrections) of the
$Z'$ mass. Therefore
the decay channel of two $W_s$'s into two
$Z'$ gauginos is forbidden as well.

Thus, we have two limits for the cross section
\begin{eqnarray}
\sigma|v| &\simeq& 16 N_{Z^\prime} \pi
\frac{M^2}{M^4_{Z^\prime}},~~~~~~~~~~~{\rm if}~ T_R<M \nonumber
\label{go1}
\eeqa
and
\beqa
\sigma|v| &\simeq& \frac{8}{3} N_{Z^\prime}
\pi \frac{{s}}{M^4_{Z^\prime}},~~~~~~~~~~~ {\rm if} ~T_R>M.
\label{go2}
\end{eqnarray}
In the non--relativistic case ($T_R<M$) we can use Eq.~(\ref{infla2})
and obtain
\begin{equation}
M> T_R\left[25+\frac{1}{2}\ln\left(\frac{M^5}{M_{Z^\prime}^4T_R}\right)\right].
\end{equation}
We now consider the relativistic case ($T_R>M$).
In this case the value we obtain
for  $\lambda$ in Eq.~(\ref{boltzmann2})
is temperature dependent.
In order to estimate Eq.~(\ref{go2}) we use the thermal average of the
energy squared
\begin{equation}
\langle{s}\rangle=4\langle E^2 \rangle
\simeq\left[\frac{5}{4}\right]40 T^2.
\end{equation}
Here the factor $[5/4]$ is only appropriate for fermions.
Using this result, it is straightforward to rewrite the Boltzmann equation
in the form
\begin{eqnarray}
\frac{dY}{dx}&=&-\frac{\langle \sigma |v| \rangle {\rm s}_e}{xH} Y^2_{EQ}
\nonumber\\ &=&-
4.3 N_{Z^\prime} g^2_{\rm eff} g_*^{-1.5}
\frac{m_{pl}M^3}{M^4_{Z^\prime}} x^{-4}
\end{eqnarray}
and get the relic density of this particle
by integrating this equation from the reheating temperature
to the present temperature:
\begin{equation}
Y_0=1.44 N_{Z^\prime} g^2_{\rm eff} g_*^{-1.5}
\frac{m_{pl}T_R^3}{M^4_{Z^\prime}}.
\end{equation}
If we insert this result into Eq.~(\ref{bound}) we obtain
\begin{equation}
 \frac{MT_R^3}{M^4_{Z^\prime}}< 6.9\times 10^{-25}
\left(\frac{10^{19}{\rm GeV}}{m_{pl}}\right)
\left(\frac{g_*}{200}\right)^{1.5} \frac{1}
{N_{Z^\prime} g^2_{\rm eff}}.
\label{ca}
\end{equation}
Since there are three unknown parameters ($M,\, M_{Z'},\, T_R$)
in Eq.~(\ref{ca}) we cannot
infer a definite value for the mass of the $W_s$ and $Z'$, but
we nevertheless deduce that $Z'$ should be very heavy.

There is still another possibility to be considered.
The W--singlet can be a triplet of a hidden
gauge group. If this gauge group is not broken, then the W--singlets
can annihilate into two hidden gauge bosons. This will lead
to the same result as for the triplet case (the uniton)
and we can obtain similar mass bounds.

\setcounter{footnote}{0}
\section{Conclusion and discussion}

In this paper we studied the cosmological constraints
on the exotic matter states that appear in the massless spectrum of
realistic free fermionic superstring models.
The free fermionic superstring models are among
the most realistic string models constructed to date,
and reproduce many of the observed properties of the Standard
Model. Among those, the replication of three and
only three families and the qualitative spectrum of
fermion masses \cite{TOP,CKM}.
The realistic nature of the free fermionic models
is perhaps not accidental but may reflect deeper properties of string
compactification, which are at present unknown. Indeed, the free
fermionic models are constructed at a highly symmetric
point in the moduli space and the appearance of three generations
is deeply rooted in the underlying $Z_2\times Z_2$ orbifold
structure \cite{foc}.

In the derivation of the Standard Model from superstring theory
we start with some larger symmetry which is subsequently
broken to the Standard Model. Absence of adjoint representations
in the massless spectrum of level one Kac--Moody algebras
restricts the possible gauge groups in the effective
low energy field theory. Furthermore, proton lifetime constraints
motivate the hypothesis that the Standard Model must be
derived directly from string theory, without an intermediate
non--Abelian gauge symmetry.  Within the free fermionic
construction the breaking is achieved by constructing boundary
condition basis vectors which are equivalent to Wilson lines
in the geometrical formulation. The use of Wilson lines
to break the non--Abelian gauge symmetries is quite generic
in superstring theory.

The breaking of the non--Abelian gauge symmetries by Wilson lines
has an important feature: it produces matter representations
that do not fit into multiplets of the original unbroken
gauge symmetry. This is an important feature as it may result
in local discrete symmetries that forbid the decay of the
``Wilsonian'' matter states into the lighter Standard Model
particles. Superstring models thus provide an intrinsic
mechanism that produces heavy stable states.

The ``Wilsonian'' matter states are classified by the
patterns of symmetry breaking, induced by the ``Wilson''
lines. In the free fermionic models the underlying $SO(10)$
gauge symmetry is broken to $SO(6)\times SO(4)$, $SU(5)\times U(1)$ or
$SU(3)\times SU(2)\times U(1)^2$.
All three cases give rise to fractionally
charged states with $Q_{\rm e.m.}$. These states may all be confined,
they may all be superheavy,
or they may diluted by inflation. Nevertheless,
it may be worthwhile to search for such states in experimental searches
for rare matter.

Of further interests are the exotic states which appear specifically
when the symmetry is broken directly to the Standard Model.
The ``Wilsonian'' sectors then contain also states which
carry the Standard Model charges, but with fractional
charge under the $U(1)_{Z^\prime}$ symmetry. We have
shown, in a specific model, that these states can, in fact,
be stable. This is an exciting observation, for then
the stable ``Wilsonian'' matter states can be natural
dark matter candidates. Furthermore, their stability
arises due to a well motivated local discrete symmetry \cite{KW}.

The superstring derived standard--like models give rise
to ``Wilsonian'' color triplets, electroweak doublets with the
standard charge assignment, and to Standard Model singlets.
Of those the Standard Model singlets are the most suited to
be the dark matter.

The appearance of a good dark matter candidate
in the superstring derived standard--like models
provides further motivation to focus on the phenomenology of
this class of models.
Although our analysis and results are limited to the
models in the free fermionic formulation,
the mechanism which gives rise to the ``Wilsonian'' matter states
is generic in superstring models.  Thus, exotic ``Wilsonian'' matter
states may be the generic, long--sought, signature of string
unification. It is of further interest to study other phenomenological
properties of these states. Such questions are currently under
investigation and will be reported in future publications.

\section*{Acknowledgments}

We thank R. Field, D. Kennedy, D. Lichtenberg, R. Oakes, P. Ramond,
P. Sikivie and C. Thorn for discussions.
Special thanks to K. Dienes for comments on the manuscript.
AEF thanks the Institute for Advanced Study for its
hospitality while part of this work was conducted.
CC thanks E. Herman, L. Gordon, A. White and the
Theory group at Argonne for hospitality.
This work was supported in part by DOE Grant
No.\ DE--FG--0586ER40272 and by KOSEF.

\section*{Appendix A. The Mass splitting}
\renewcommand{\thesection}{{A}}

In this appendix we comment on the mass splitting
relation between the two
heavy--light mesons $U_0$ and $U_{-1}$.
The goal of our discussion here is
to show in a semi-quantitative way
that the mass splitting between these two heavy-light mesons is
compatible with zero and even with a $U_0$ lighter than $U_1$.

While transition and elastic form factors are
calculable in the heavy quark effective theory,
the calculation of the static properties of hadrons, such as the
electromagnetic mass splitting of mesons
containing heavy--light quarks, relies
 on drastic approximations with a built--in model dependence.

However, a simple estimate of the
mass splitting between the two states $U_0$
$(\bar{D}d)$ and $U_{-1}$ (${D}\bar{u}$) can be obtained in the context of
 potential models within a non--relativistic approximation.
Although this may not look appropriate,
it has been shown that, even in the context of light mesons, the
non--relativistic approximation to the analysis
of the spectra of these systems
(which are relativistic)
has been quite successful when a Coulomb $+$
linear $+$ constant potential
has been used to model the interaction. On the other hand, relativistic
approaches based on bag models \cite{thorn}
have also produced satisfactory
results. In ref. \cite{rose} it has been argued that
there is an interesting duality relation between the
bag model solutions (with a quadratic potential) and
those produced by non--relativistic potential models.
This observation seems to clarify, to some extent,
that the non--relativistic
approach is in part consistent even for systems which are intrinsically
relativistic \cite{flam}. With these motivations in mind,
let us first analyze the electromagnetic (plus mass difference of the
two constituent quarks) contributions to the mass
splitting.
The interaction may be taken to be Coulomb (at short distances) + linear
(the confining tail).
We may assume that, in a meson, when one of the
two quarks gets very heavy,
the bound state is hydrogen--like.
If we neglect the effect of the strong interactions (which is
a questionable assumption
since $\alpha_s$ is not small at distances of 1 $fm$)
the potential of the
Breit--Fermi Hamiltonian corresponding to
the electromagnetic interaction,  for
states of with zero angular momentum is given by

\beq
V_{em}= \alpha_{em}{Q_1 Q_2\over r} -
\alpha_{em}{ 8\pi\over 3}{{\bf S}_1\cdot {\bf S_2}\over m_1 m_2}
\delta^{(3)}({\bf r}).
\eeq
$m_1$ and $m_2$ are the masses of the two
quarks and $Q_i$ the corresponding
electromagnetic charges.
For spin singlet configurations
\beq
\langle V\rangle_{0^-}=\alpha_{em}\langle
Q_1 Q_2\rangle\left(\langle{1\over r}\rangle +
{1\over m_1 m_2} |R(0)|^2\right),
\eeq
with $R(r)$ the radial wave function of the
ground state ($\psi(r)=(1/\sqrt{4
\pi})R(r)$). The spin interactions drops
out in the heavy mass limit of
any of the two quarks.
Under these assumptions we get
 \beq
M_{U_0}-M_{U_{-1}}= (m_d - m_u) -
{\alpha_{em}\over 3}\langle{1\over r}\rangle.
\eeq
If we use the estimate $m_d - m_u\approx M_{K^0}-M_{K^+}=4$ MeV for the
constituent mass of the two light quarks
\cite{lucha} and a size of the bound state of $1 fm$, then we get that
the electromagnetic corrections are $1$ MeV, thereby giving
a positive $M_{U_0}-M_{U_{-1}}\approx 3$ MeV.
This difference can be reduced
by strong interaction effects which can provide a
$\Delta M_s$ which can even
over--compensate $\Delta M_{em}$ \cite{lucha}.
In fact,
the difference between the
up and the down quark masses causes a
difference between the expectation values
of the strong interaction Breit--Fermi Hamiltonian.
This effect, therefore,
can be responsible of a reduction of the mass
splitting between the two states.
We remark here that, in the case of neutral
and charged B mesons, all these
arguments do apply. The particle data book \cite{data}
gives a mass splitting between
$B^0$ and $B^{-}$ which is compatible with zero.

Let us now comment on the estimate of the electromagnetic
interaction to the mass splitting
of heavy--light mesons in the context of the
heavy quark effective theory.
At the same time we will also comment
on the strong interaction contribution
to the splitting
which plays a significant role --but is still not calculable--
in the context of these same effective theories.
Unfortunately, the evaluation of the binding energy of the light
degrees of freedom from the heavy quark theory suffers from a large
uncertainty. The discussion which follows
has been included to make this point
more explicit.

In the heavy quark limit, the interaction between the quark
spin and the chromomagnetic field vanishes, being inversely proportional
to the heavy quark mass $m_Q$. Beside the spin symmetry,
a new symmetry appears
in the effective description of the bound state,
since the interaction between
the heavy
and the light degrees of freedom ($u,\, d$) is flavor blind.
We picture the heavy quark then acting as a static color force.
There are two basic scales appearing in the effective theory: $m_Q$ and
$\bar{\Lambda}/m_Q\equiv\,(m_{hadron} - m_Q)/m_Q$, with $\bar{\Lambda}$
the scale of the light degrees of freedom.
While $m_Q$ sets the scale for the
perturbative expansion (in $\alpha_s(m_Q)$)
of the theory, the second scale
characterizes the nonperturbative contributions
of additional form factors
in the form of matrix elements of higher dimensional operators.
Notice that the QED analogy breaks down exactly
due to the presence of this second scale.
To leading order in $1/m_Q$,
$\bar{\Lambda}=m_{hadron} - m_Q $ is flavor and
spin independent. From QCD sum rules, using the pole mass for $m_Q$
(and neglecting renormalon ambiguities (see \cite{mannel}
and refs. therein) one gets the estimate

\beq
\bar{\Lambda}=570 \pm 70 {\rm MeV}
\label{lambda}
\eeq
(see \cite{mannel} and refs. therein).
The uncertainty in Eq.~(\ref{lambda}) largely over compensates the
electromagnetic contribution to the splitting.

Let us now comment briefly on the electromagnetic breaking contributions
as estimated in the context of the heavy quark effective theory in ref.
 \cite{LS}.
The calculation of the electromagnetic splitting
has been based, in the past,
on the use of dispersion relations for forward Compton scattering
\cite{Feynman}.
We remark that the study of factorization in exclusive processes
\cite{BL} \cite{BS} and of
dispersion relations, in particular for fixed angle Compton
scattering \cite{cc}, has a long and involved history.
In the forward region, where perturbative QCD breaks down, the treatment
of the process suffers from a strong model dependence.
Therefore, the
use of dispersion relations at very low momentum transfer, as needed
in the case of the calculation of mass splittings,
is an uncharted territory.
The basic strategy of the method in \cite{LS}
is to relate the order $e^2$ isospin breaking
corrections to the forward Compton
scattering amplitude $T$ and then use the
heavy quark effective theory
in the large $N_c$ limit to write $T$ in
terms of heavy meson form factors.

Neglecting $1/m_q^2$ corrections, the authors of ref. \cite{LS} give
\beq
M_{U^+}-M_{U^0}\sim +1.7 -0.13\left({\beta\over 1 {\rm GeV}^{-1}}\right) -
0.03\left({\beta\over 1 {\rm GeV}^{-1}}\right)^2,
\eeq
where $\beta\sim 1/m_Q$ measures the matrix element of the decay of the
first excited heavy meson state into the ground state plus a photon.

The use of large $N_c$ arguments and the ansatze used to match
the dimensional counting behavior of the heavy quark form factor
introduce a model dependence into the theory.
The authors of ref. \cite{LS} argue that their
results are accurate within $30\%$. Notice that this result is
compatible with the naive estimate
obtained from the quark model discussion we have presented above.
Also, notice that the combination of this result with Eq. (\ref{lambda}),
in particular the large uncertainty in the estimate of
$\Lambda$ tells us that $M_{U_0}-M_{U_{-1}}$
is not incompatible with zero,
as suggested by
potential model calculations.

\section*{Appendix B.  Quark Decay Modes}
\renewcommand{\thesection}{{B}}
\setcounter{equation}{0}
In this appendix we summarize the derivation of the Feynman rules for
$N=1$ supersymmetric QCD and define our
conventions for the various diagrams.

The Lagrangian is defined by \cite{HK}
\begin{equation}
{\cal L}=\bar{q}i\!\dnot{D}q -\frac{1}{4}F^{\mu\nu}F_{\mu\nu}
+(D^{\mu}\tilde{q})^*D_{\mu}\tilde{q}
\end{equation}
where
\begin{equation}
\dnot{D}_{ij}= \dnot{\partial}\delta_{ij}-\frac{ig}{2}\!\dnot{A}_aT^a_{ij}
\end{equation}
In SUSY QCD
there is a gluino--gluino--gluon term,
\begin{equation}
{\cal L}_{g\tilde{g}\tilde{g}}=\frac{i}{2}g_s f_{abc}\bar{\tilde{g}}_a
\gamma_\mu\tilde{g}_bA^\mu_c~;
\end{equation}
quark--squark--gluino term.
\begin{equation}
{\cal L}_{q\tilde{q}\tilde{g}}=-\sqrt{2}g_s T^a_{jk}\sum_{i=u,d}
\left(\bar{\tilde{g}}_a P_L q^k_i\tilde{q}^{j*}_{iL} +
\bar{q}^j_{i} P_R \tilde{g}_a \tilde{q}^k_{iL} -
\bar{\tilde{g}}_a P_R q^k_{i} \tilde{q}^{j*}_{iR} -
\bar{q}^j_{i} P_L \tilde{g}_a \tilde{q}^k_{iR} \right),
\end{equation}
where $i$ is summed over the three families of
quarks and $T^a=\lambda^a/2$
are the usual $SU(3)$ generators.
The two chirality projectors are defined by
$P_L=(1 - \gamma_5)/2$ and $P_R=(1+\gamma_5)/2$.
The generators of $SU(3)$ are taken to be hermitian.
The quark--gluon vertex, in these conventions, is given by

\beq
{\cal L}{q \bar{q} g}=-\bar{q}A_{\mu}^a \gamma^\mu q T^a.
\eeq
In momentum space we get $-i g_s \gamma_\mu T^a$
for the $q\bar{q} g$ vertex
and $-g_s f_{a b c}\gamma_\mu$ for the $\sg \sg g$ vertex.

We introduce the following expansions for the Dirac fermions (quarks),
Majorana fermions (gluinos) and for the scalar quarks

\beq
q(x)=\int{d^3 k\over (2 \pi)^3}\left({m\over k_0}\right)
\left (b(k)u(k)e^{-i kx} + d^\dagger(k) v(k) e^{i k x}
\right)
\label{one}
\eeq
\beq
\sg(x)=\int{d^3 k\over (2 \pi)^3}\left({m\over k_0}\right)
\left(\sbb (k)\tilde{u}(k)e^{-i kx} + \sbb^\dagger(k)\tilde{ v}(k) e^{i k x}
\right)
\label{two}
\eeq

\beq
\sq(x)=\int {d^3 k\over (2 \pi)^3 2 \omega_k}
\left( a(k)e^{-i kx} + b^\dagger (k) e^{i kx}\right).~~~~~~~~~~~~~
\label{three}
\eeq
Notice that in (\ref{one}) (\ref{two}) and (\ref{three})
we have omitted the
sum over the polarizations for simplicity.
We enforce the usual quantization conditions
\beqa
&& \left\{b(k),b^\dagger (k')\right\}=
(2\pi)^3\left({k_0\over m}\right)
\delta^3({\bf k}-{\bf k}'); \nonumber \\
&& \left\{\sbb(k),\sbb^\dagger (k')\right\}=
(2\pi)^3\left({k_0\over m}\right)
\delta^3({\bf k}-{\bf k}'); \nonumber \\
&& \left\{d(k),d^\dagger (k')\right\}=
(2\pi)^3\left({k_0\over m}\right)
\delta^3({\bf k}-{\bf k}').
\eeqa

\begin{figure}
\centerline{\epsfbox{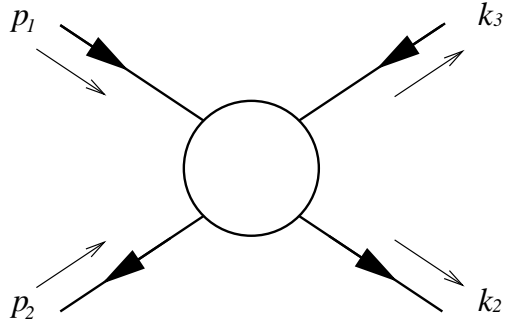}}\label{tree}
\caption{Momentum and fermion number flow}
\end{figure}
Scattering amplitudes can be defined in terms of the vacuum--to--vacuum
correlators either by using the LSZ reduction formula or, more simply, by
calculating
expectation values of the interaction terms in initial and final states.
\beq
S_{fi}= {(ig_s)^2\over 2}\int d^4 y_1 d^4 y_2
\langle Q\bar{Q}|{\cal L}(y_1){\cal L}(y_2)|\sg\sg\rangle .
\eeq
We have defined
\beq
|Q\bar{Q}\rangle\equiv b^\dagger
(p_1)d^+(p_2)|0\rangle;\,\,\,\,\,\,\,\,\,\,\,
|\sg\sg\rangle =\tilde{b}^\dagger (k_3)\tilde{b}^\dagger (k_2)|0\rangle .
\eeq
Notice that it is convenient to fix a specific
convention by labeling one of the two gluinos as first particle $(k_2)$
and the second one as second particle $(k_3)$.
Define $(A B)_c\equiv \langle 0| A B|0\rangle $ to be the usual Wick
contraction and set (symbolically)
\beqa
&& q(x)\sim b u + d^\dagger \bar{v}\nonumber \\
&& \bar{q}(x)\sim b^\dagger \bar{u} + d \bar{v} \nonumber \\
&& \sg (x)\sim \tilde{b} u +\tilde{b}^\dagger \tilde{v}\nonumber \\
&& \bar{\sg} (x)\sim \tilde{b}^\dagger \tilde{ u} +
\tilde{b} \bar{\tilde{v}}\nonumber \\
&& \sq (x) \sim a + a^\dagger,
\eeqa
in order to get the only non--vanishing
contractions with the initial (final)
states
\beqa
&& (q(x) b^\dagger (k))_c=u(k) e^{-i k x}\nonumber \\
&& (\bar{q}(x) d^\dagger (k))_c= \bar{v}(k)e^{-i k x}\nonumber \\
&& (\sg(x)\sbb^\dagger (k))_c=\tilde{u}(k)e^{-i k x}\nonumber \\
&& (\bar{\sg}(x)\sbb^\dagger (k))_c=
	\bar{\tilde{v}}(k) e^{-i k x}\nonumber \\
&& (b(k) \bar{\sg}(x))_c=\bar{\tilde{u}}(k)e^{i k x}\nonumber \\
&& (b(k) \sg (x))_c=\tilde v(k)e^{-i k x}.\nonumber \\
\eeqa
\begin{figure}
\centerline{\epsfbox{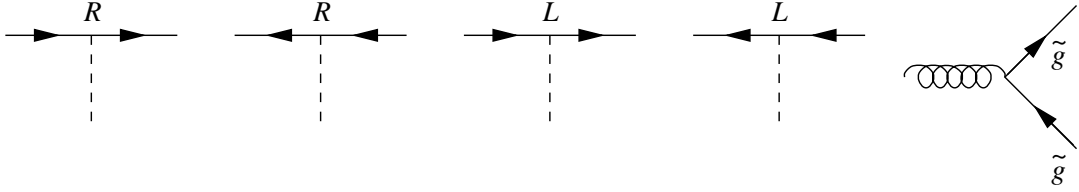}}\label{vertices}
\caption{Gluino vertices}
\end{figure}

There are 5 diagrams which contribute in the supersymmetric case to the
annihilation of a $Q\bar{Q}$ pair into two gluinos.
In our calculation we neglect mixing between the squark fields.
The amplitudes are given by
\beqa
&&A_1 =-{2 i \alpha_s\over t- M_s^2}
	(T^b T^a)_{jk}\bar{\tilde{u}}^a(k_3)P_R
u^k(p_1)\bar{v}^j(p_2)P_L\tilde{v}^b(k_2) \nonumber \\
&& A_2= {2 i \alpha_s\over t- M_s^2 }
	(T^b T^a)_{jk}\bar{\tilde{u}}^a(k_2)P_R
u^k(p_1)\bar{v}^j(p_2)P_L\tilde{v}^b(k_3) \nonumber \\
&&A_3 =-{2 i\alpha_s\over t- M_s^2}
	(T^b T^a)_{jk}\bar{\tilde{u}}^a(k_3)P_L
u^k(p_1)\bar{v}^j(p_2)P_R\tilde{v}^b(k_2) \nonumber \\
&& A_4= {2 i\alpha_s\over t - M_s^2}
	(T^b T^a)_{jk}\bar{\tilde{u}}^a(k_2)P_L
u^k(p_1)\bar{v}^j(p_2)P_R\tilde{v}^b(k_3) \nonumber \\
&& A_5=-{1\over 2 \,\,s}\alpha_s T^d_{j k}f^{d a b}
\bar{\tilde{u}}^a(k_3)\gamma^\mu
\tilde{v}^k(k_2))\bar{v}^j(p_2)\gamma_{\mu}u^k(p_1) \nonumber \\
\eeqa
where we have set $s=(p_1 + p_2)^2$ and $t=(p_1 - k_3)^2$.
Here $a$ and $b$ are SU(3) octet indices for the two gluino spinors
$\tilde{v}$ and $\tilde{u}$, while $j$ and $k$
are the color indices in the
fundamental. We have assumed that the left and right squarks have equal
masses ($M_s$). We also take the final gluinos to be massless.
Notice that in the case of annihilation into a quark--antiquark pair, the
number
of diagrams is doubled (with a factor of
two also for the $A_5$, absent in the
QCD case)
because of additional non--zero contractions with the
Majorana final states.  Notice that in the direct and exchange diagrams
we have distinguished between the fermion flow and the momentum flow.
In each exchanged $t$--channel diagram
the fermion flow for the final states
is reversed compared to the flow for each direct diagram.

\begin{figure}
\centerline{\epsfbox{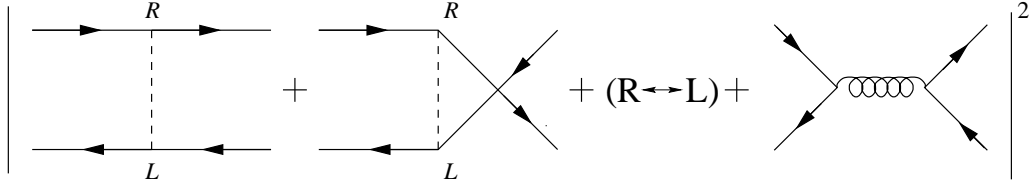}}\label{u2sg}
\caption{$Q\bar{Q}\longrightarrow \tilde{g}\tilde{g}$ annihilation}
\end{figure}
The averaged cross section for the annihilation of
the uniton into two gluinos or into
two gluons is generically defined by
\beq
 {\langle |{\cal M}_U|^2\rangle}= {1\over (3)^2 (2)^2}|{\cal M}_U|^2
\eeq
where we have averaged over spin and color.
The calculation has been performed in the
most general case, taking all the mass
parameters to be different.
The final result is too cumbersome to be given here.

For the annihilation of two heavy quarks into two gluinos in
the equal--mass limit we get

\begin{eqnarray}
&& |{\cal M}|^2=\frac{256\pi^2\alpha_s^2}
   {27s^2( t-{m^2}  ) (  s + t -{m^2} ) }
( 9m^8 - 18m^6s + 13m^4s^2 - 8m^2s^3 \nonumber \\
&& +(13s^3- 44m^2s^2 + 63m^4s -
       36m^6 )t \nonumber \\
&& + ( 31s^2- 72m^2s + 54m^4)t^2 +(27s- 36m^2t^3 )t^3 + 9t^4).
\end{eqnarray}
The total cross section is
\begin{eqnarray}
\sigma= \frac{16\pi\alpha^2_s} {27A^3}\left(-24m^4 - 22m^2s + 7s^2
-m^2\log\left(\frac{s+A}{s-A}\right) \right).
\end{eqnarray}
We have set $A=\sqrt{s(s-4m^2)}$. In the non--relativistic limit we get
$s=4m^2(1+v_{cm}^2)$ and $A=4m^2 v_{cm}$ with a total cross section
\begin{equation}
\sigma|v|={{64\pi\alpha^2_s }\over {27m^2}}
+ {{\rm O}(v^2)}
\end{equation}
where $|v|=2|v_{cm}|$

Let us now comment on the  $Q \bar{Q}$ annihilation into two gluons.
The calculation is performed in the
Feynman gauge and we have included the ghost
contribution diagram (since there are
two gluons in the final state) to remove the unphysical polarizations.
We recall that in the case of massive
quarks all the interference diagrams
give a non--vanishing contribution.

\begin{figure}
\centerline{\epsfbox{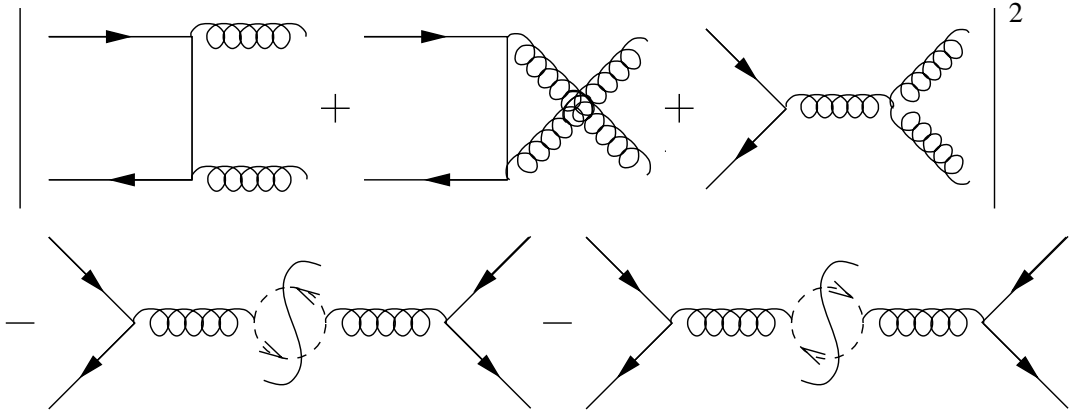}}\label{u2g}
\caption{$Q\bar{Q}\longrightarrow gg$ annihilation}
\end{figure}

We give
expressions for all the cross sections in the relativistic and in
the non--relativistic cases.
The total cross section is given by
\begin{eqnarray}
\sigma =
\frac{16\pi\alpha_s^2}{27 m^2s^2 A^2}
	\left( -(7s+31m^2)A+4(s^2+4m^2s+m^4)\log
\left(\frac{s+A}{s-A}\right) \right).
\end{eqnarray}
In the non--relativistic limit we get
\begin{equation}
\sigma|v|={{14\pi\alpha_s^2 }\over {27{m^2}}}
+ {{\rm O}(v^2)}.
\end{equation}

\begin{figure}
\centerline{\epsfbox{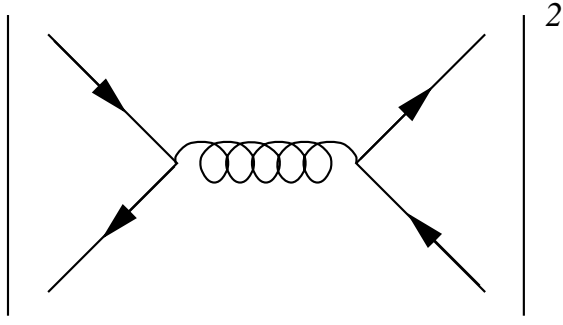}}\label{u2f}
\caption{$Q\bar{Q}\longrightarrow q\bar{q}$ annihilation}
\end{figure}
In the case of the annihilation of a heavy $Q\,\bar{Q}$
pair into a light $q\bar{q}$
pair we get

\begin{eqnarray}
\sigma= \frac{8\alpha_s^2\pi(s+2m^2)}{27s \sqrt{s(s-4m^2)}},
\end{eqnarray}
which in the non--relativistic limit gives
\begin{equation}
\sigma|v|=\frac{2\pi\alpha^2_s}{9m^2}
+ {{\rm O}(v^2)}.
\end{equation}

If we sum over all 6 quark flavors of the final state we get
\begin{equation}
\sigma|v|=\frac{4\pi\alpha^2_s}{3m^2}
+ {{\rm O}(v^2)}.
\end{equation}

\begin{figure}
\centerline{\epsfbox{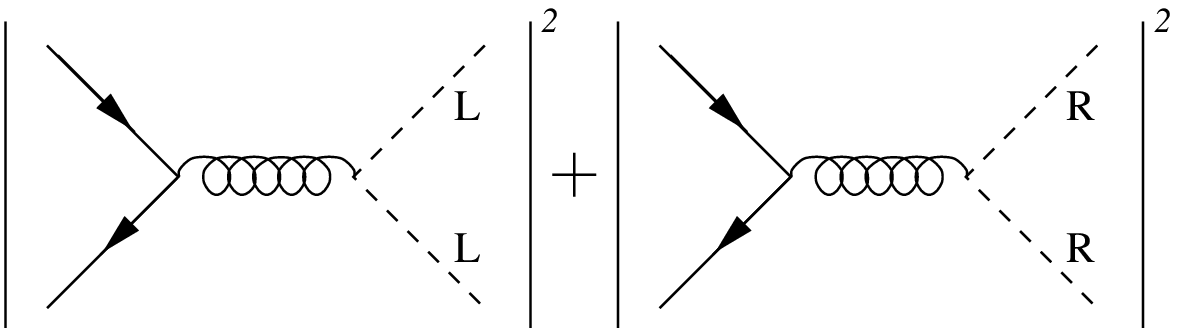}}\label{u2sf}
\caption{$Q\bar{Q}\longrightarrow \tilde{q}\tilde{q}^*$ annihilation}
\end{figure}

The total cross section for squark production from
$Q\bar{Q}$ annihilation
is given by
\begin{eqnarray}
\sigma&=& \frac{4\alpha_s^2\pi(s+2m^2)}{27s \sqrt{s(s-4m^2)}}
\nonumber\\ &=& \frac{4\alpha_s^2\pi(s+2m^2)}{27s A}.
\end{eqnarray}
In the non--relativistic limit we get
\begin{equation}
\sigma|v|=\frac{\pi\alpha^2_s}{9m^2}
+ {{\rm O}(v^2)}
\end{equation}
If we sum over all 6 flavors of the quarks in the final state we get
\begin{equation}
\sigma|v|=\frac{2\pi\alpha^2_s}{3m^2}
+ {{\rm O}(v^2)}
\end{equation}

%=========================================================================
%======================== REFERENCES =====================================
%=========================================================================

\vfill\eject

\bigskip
\medskip

\bibliographystyle{unsrt}

\end{document}